\documentclass[aps,prl,reprint,floatfix,superscriptaddress]{revtex4-2}
\usepackage[T1]{fontenc}
\usepackage[utf8]{inputenc}
\usepackage{graphicx}
\usepackage{xcolor}
\usepackage{txfonts}
\usepackage{times}
\usepackage{hyperref}
\usepackage{orcidlink}
\usepackage{enumitem}   
\usepackage[normalem]{ulem}

\definecolor{myblue}{RGB}{0, 40, 140}

\hypersetup{
    unicode=true,          
    pdftitle={Accurate Determination of Blackbody Radiation Shifts in a Strontium Molecular Lattice Clock},    
    pdfauthor={},     
    colorlinks=true,       
    linkcolor=myblue,          
    citecolor=myblue,        
    filecolor=myblue,      
    urlcolor=myblue
}

\begin{document}
	\title{Accurate Determination of Blackbody Radiation Shifts in a Strontium Molecular Lattice Clock}

    \author{B. Iritani\orcidlink{0000-0002-7911-2755}}
        \altaffiliation{These authors contributed equally to this work.}
        \affiliation{Department of Physics, Columbia University, 538 West 120th Street, New York, NY 10027-5255, USA}
        
    \author{E. Tiberi\orcidlink{0000-0001-7168-7194}}
        \altaffiliation{These authors contributed equally to this work.}
        \affiliation{Department of Physics, Columbia University, 538 West 120th Street, New York, NY 10027-5255, USA}

    \author{W. Skomorowski\orcidlink{0000-0002-0364-435X}}
    \affiliation{Centre of New Technologies, University of Warsaw, Banacha 2c, 02-097 Warsaw, Poland}

    \author{R. Moszynski\orcidlink{0009-0008-7669-3751}}
        \affiliation{Quantum Chemistry Laboratory, Department of Chemistry,
        University of Warsaw, Pasteura 1, 02-093 Warsaw, Poland}
        
     \author{M. Borkowski\orcidlink{0000-0003-0236-8100}}
        \email{mateusz@cold-molecules.com}   
        \affiliation{Department of Physics, Columbia University, 538 West 120th Street, New York, NY 10027-5255, USA}
        \affiliation{Van der Waals-Zeeman Institute, Institute of Physics, University of Amsterdam, Science Park 904, 1098 XH Amsterdam, The Netherlands}
        \affiliation{Institute of Physics, Faculty of Physics, Astronomy and Informatics, Nicolaus Copernicus University, Grudziadzka 5, 87-100 Torun, Poland}
        
    \author{T. Zelevinsky\orcidlink{0000-0003-3682-4901}}
        \email{tanya.zelevinsky@columbia.edu}
        \affiliation{Department of Physics, Columbia University, 538 West 120th Street, New York, NY 10027-5255, USA}

	\date{\today}
	
    \begin{abstract}
    Molecular lattice clocks enable the search for new physics, such as fifth forces or temporal variations of fundamental constants, in a manner complementary to atomic clocks.    
    Blackbody radiation (BBR) is a major contributor to the systematic error budget of conventional atomic clocks and is notoriously difficult to characterize and control.     
    Here, we combine infrared Stark-shift spectroscopy in a molecular lattice clock and modern quantum chemistry methods to characterize the polarizabilities of the Sr$_2$ molecule from dc to infrared. 
    Using this description, we determine the static and dynamic blackbody radiation shifts for all possible vibrational clock transitions to the $10^{-16}$ level.
    This constitutes an important step towards mHz-level molecular spectroscopy in Sr$_2$, and provides a framework for evaluating BBR shifts in other homonuclear molecules.
    \end{abstract}
    \maketitle
 
Frequency standards are the cornerstone of precision measurement. Optical atomic clocks set records in both precision and accuracy, and are poised to redefine the second \cite{Takamoto2005,Katori2011, McGrew2018, Bothwell2019, McGrew2019, Lodewyck_2019, BIZE2019153}. There is a growing interest in precision measurements with molecules~\cite{Borkowski2018, Kondov2019a, Kobayashi2019a, Hanneke2021, Barontini2022}. The simple structure of homonuclear diatoms like Sr$_{2}$ makes them ideal testbeds to probe new physics, including searching for corrections to gravity at short distances \cite{Salumbides2014, biesheuvel2016probing, Borkowski2019, Heacock2021} and temporal variation of fundamental constants \cite{Schiller2005,Demille2008,Beloy2011,Schiller2014,Kajita2014, Germann2014, Wcislo2018,Safronova2019,Hutzler2020,Lange2021, Barontini2022}. Thus, there is interest in improving techniques for molecular spectroscopy. Even for ultra-precise atomic clocks, the blackbody radiation (BBR) shift remains a primary contribution to the uncertainty of the clock measurement \cite{LeTargat2013, Falke2014, Nicholson2015a, Koller2017, McGrew2018, Bothwell2019, Hisai2021, Ohmae2021}, and is notoriously difficult to control and characterize \cite{Ushijima2015, Ablewski2020, Yudin2021}. State-of-the-art evaluations of BBR shift rely on measurements of the differential dc polarizability of the clock states in conjunction with modeling of dynamic contributions~\cite{Middelmann2011,Middelmann2012,Lisdat2021, Porsev2006,Safronova2013}.

\begin{figure}[t]
  \centering
  \includegraphics[width = 0.95\columnwidth]{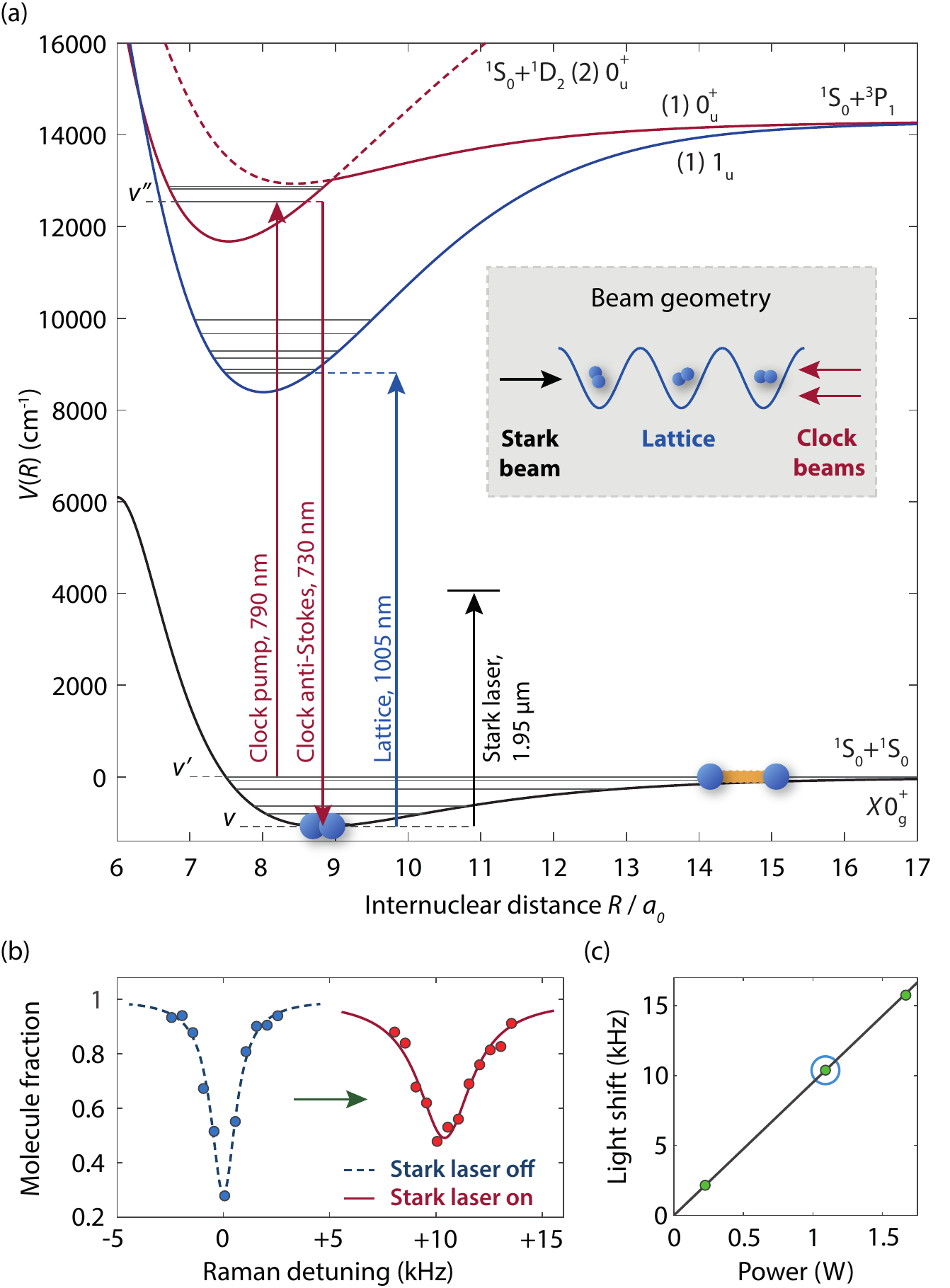}
    \caption{Stark-shift spectroscopy in Sr$_2$ on the example $0\leftrightarrow62$ transition. (a) We rely on narrow two-photon Raman transition via an intermediate state in the $(1)\,0_u^+$ (red arrows) in a magic lattice that couples the deeply-bound clock state $v$ to an excited $(1)\,1_u$ state (blue arrow). (b) We induce Stark shifts to probe differential polarizabilities of ground ro-vibrational states with $1.95\,$\textmu{}m{} light. (c) Example light shift measurement. The encircled point corresponds to subplot~(b). }
    \label{fig:cartoon}
 \end{figure}

Previously, we demonstrated record precision and accuracy for a molecular lattice clock by measuring a 32-THz transition between two vibrational levels in ultracold Sr$_2$ molecules, reaching a $4.6\times10^{-14}$ systematic uncertainty~\cite{Leung2023}. Estimates of the BBR contribution to this uncertainty relied on preliminary theoretical modelling of polarizabilities that lacked experimental verification. Here, we determine room-temperature BBR shifts for our molecular clock to the 10$^{-16}$ level. To do so, we employ modern quantum chemistry methods to determine the differential polarizabilities for all vibrational clock transitions and verify our theory directly by measuring Stark shifts induced by a mid-infrared laser for a wide variety of molecular clock transitions (Fig.~\ref{fig:cartoon}). Given this combined experimental and theoretical picture, we develop a complete description of the BBR effect for all vibrational levels within the ground-state potential of $^{88}$Sr$_2$ molecules.

The experimental sequence closely follows that of our previous works~\cite{Kondov2019a, Leung2020, Leung2021, Leung2023}. A~$2$-\textmu{}K sample of ultracold strontium atoms is trapped in a one-dimensional, horizontal, near-infrared optical lattice. We form weakly bound molecules via a photoassociation pulse tuned to the \mbox{$-353$-MHz} $1_u$ resonance~\cite{Zelevinsky2006}. This bound state predominantly decays to a pair of rotational $J=0, 2$ states of the top vibrational state, $v'=62$, in the ground-state potential. We then apply a two-photon Raman pulse to probe selected clock transitions. We detect the number of remaining $v'=62$ molecules by photodissociation~\cite{McGuyer2015a} and counting the recovered atoms. Unless otherwise specified, we always refer to rotationless $J=0$ states in the electronic ground-state potential, and list the lower energy state first for a given transition, regardless of where the molecular population is initialized.

We rely on narrow-linewidth Raman transitions between the least bound $v'=62$ vibrational state and selected deeply bound vibrational states $v$ [Fig.~\ref{fig:cartoon}(a)]. We address each of these transitions via intermediate states $v''$ in the electronically excited $(1)\,0_u^+$ potential. The vibrational splittings are determined by the difference in the pump ($v'\rightarrow v''$) and anti-Stokes ($v''\rightarrow v$) laser frequencies. We select intermediate states with favorable Franck-Condon factors for the pump and anti-Stokes transitions for each interrogated pair of clock states (Table~\ref{tab:results}). We address clock states throughout the potential well using three different intermediate states in the excited $(1)\,0_u^+$ potential: $v''~=~11$ [at $-57\,084\,156.51(12)$~MHz from the $^1$S$_0$+$^3$P$_1$ threshold], $v''~=~15$ [at $-48\,855\,512.13(18)$~MHz], and $v''~=~16$ [at $-47\,036\,433.95(23)$~MHz]. The selection of intermediate states is a balancing act between available lasers and transition strengths, and required several diode lasers in the 727--735~nm and 760--800~nm wavelength ranges. 

We locate the vibrational states $v$ using Autler-Townes spectroscopy: we first induce molecular loss with the pump laser, and then scan the anti-Stokes laser until the line is split into a doublet \cite{autler1955, townes2013microwave, Jones2006, martinez2008, Kitagawa2008, Leung2021}. While high-precision absolute determinations of these binding energies are beyond the scope of this Letter, we list the vibrational splittings $f_{v \leftrightarrow v'}$ to $<$100 kHz (Table~\ref{tab:results}). The uncertainty is fully dominated by light shifts (Supplemental Material).

\begin{table*}
  \caption{Investigated $^{88}$Sr$_2$ molecular states. The initial state is always the rotationless top $v'=62$ level; $v$ denotes the target level in the $^1$S$_0$+$^1$S$_0$ $0_g^+$ ground state and $\lambda_{\rm magic}$ is the magic wavelength. The differential polarizabilities are expressed in atomic units of $e^2 a_0^2 / E_h$, where $e$ is the electron charge, $a_0$ is the Bohr radius and $E_h$ is the Hartree energy~\cite{Tiesinga2021}. The error bars on theoretical polarizabilities stem from comparison to experiment.}
  \begin{ruledtabular}
  \begin{tabular}{rrrrcrrrc}
    \multicolumn{3}{c}{Clock transitions} & & 
    & \multicolumn{3}{c}{Differential polarizability $\alpha_{v\leftrightarrow v'}(\omega)$ (a.u.)} \\
    \cline{1-4} \cline{6-8}
    $X\,0_g^+$ $v\leftrightarrow v'$ & $v''$  & $f_{v\leftrightarrow{}v'}$ (MHz) & $\tilde R_v$ (a.u.) &  $\lambda_{\rm magic}$ (nm) & Exp. ($1.95\,$\textmu{}m) & Th. ($1.95\,$\textmu{}m) & Th. (dc) & $\Delta f_{v\leftrightarrow v'}$ (Hz)\\
    \colrule
    $61\leftrightarrow 62$ & 15 & 1263.673\,58(20)\,\cite{McGuyer2015a} & 43.6 & -- & $-0.41(0.52)$ & $-0.1326(35)$ & $-0.1080(28)$ &  $+9.32(25) \times 10^{-4}$\\
    $55\leftrightarrow 62$ & 15 & 108\,214.221(10) & 21.6 &  -- & $-3.68(0.38)$ & $-2.985(78)$ & $-2.429(63)$ & $+0.020\,99(56)$\\
    $41\leftrightarrow 62$ & 11 & 2\,177\,876.735(81) & 13.6 & 996.4379(10) & $-21.67(0.88)$ & $-19.10(50)$ & $-15.60(41)$ & $+0.134\,9(37)$\\
    $27\leftrightarrow 62$ & 11 & 8\,075\,406.280(18) & 11.1 & 1006.5787(10) & $-40.4(1.8)$ & $-39.3(1.0)$ & $-31.99(84)$ & $+0.276\,8(75)$\\
    $12\leftrightarrow 62$ & 16 & 19\,176\,451.651(35) & 9.62 & 1007.7634(10) & $-60.1(4.0)$ & $-61.3(1.6)$ & $-49.7(1.3)$ & $+0.430(12)$\\
    $7\leftrightarrow 62$ & 15 & 24\,031\,492.422(24) & 9.27 &   1007.1334(10) & $-66.0(2.5)$ & $-68.3(1.8)$ & $-55.1(1.4)$ & $+0.477(13)$\\
    $1\leftrightarrow 62$ & 11 & 30\,640\,159.753(75) & 8.91 &   1016.9714(10) & $-75.7(3.3)$ & $-76.0(2.0)$ & $-61.1(1.6)$ & $+0.529(15)$\\
    $0\leftrightarrow 62$ & 11 & 31\,825\,183.207\,5928(51)\,\cite{Leung2023} & 8.86 &   1004.7720(10) & $-76.4(3.6)$ & $-77.2(2.0)$ & $-62.1(1.7)$ & $+0.538(15)$\\
  \end{tabular}
  \end{ruledtabular}
  \label{tab:results}
\end{table*}

By employing several strategies to achieve 1-kHz spectroscopic resolution, we can determine ac Stark shifts to $\sim$150\,Hz using Lorentzian fits (Supplemental Material). After initially locating the transitions, we switch to a Raman configuration by detuning $+30$~MHz from the intermediate resonance to narrow down our transition linewidth. We stabilize the pump laser to a high finesse (>$3\times10^5$) cavity using a Pound-Drever-Hall lock~\cite{Pound1946, Drever1983}, which in turn provides a stable reference for the repetition rate of an optical frequency comb. We then lock our anti-Stokes clock laser to the frequency comb. This locking scheme ensures the stability of the frequency difference between the two Raman lasers. In addition to stabilizing our clock lasers, we rely on magic trapping to reduce inhomogenous broadening. Our method utilizes polarizability crossings generated by the dispersive behavior of the target state polarizability near transitions to the electronically excited $(1)\,1_u$ potential \cite{Kondov2019a}. We select $(1)\,1_u$ states such that the line strength $S$ \cite{Leung2020} is greater than $\sim10^{-5}$~$(ea_0)^2$ (here $e$ is the electron charge, $a_0$ is the Bohr radius). Large line strengths correspond to large magic detunings, allowing few-ms molecular lifetimes, and Fourier-limited linewidths of 1~kHz or better. Our lattice laser is wavemeter-locked to $\sim$30\,MHz precision. 

\begin{figure}[b]
\centering
\includegraphics[width=0.95\columnwidth, clip]{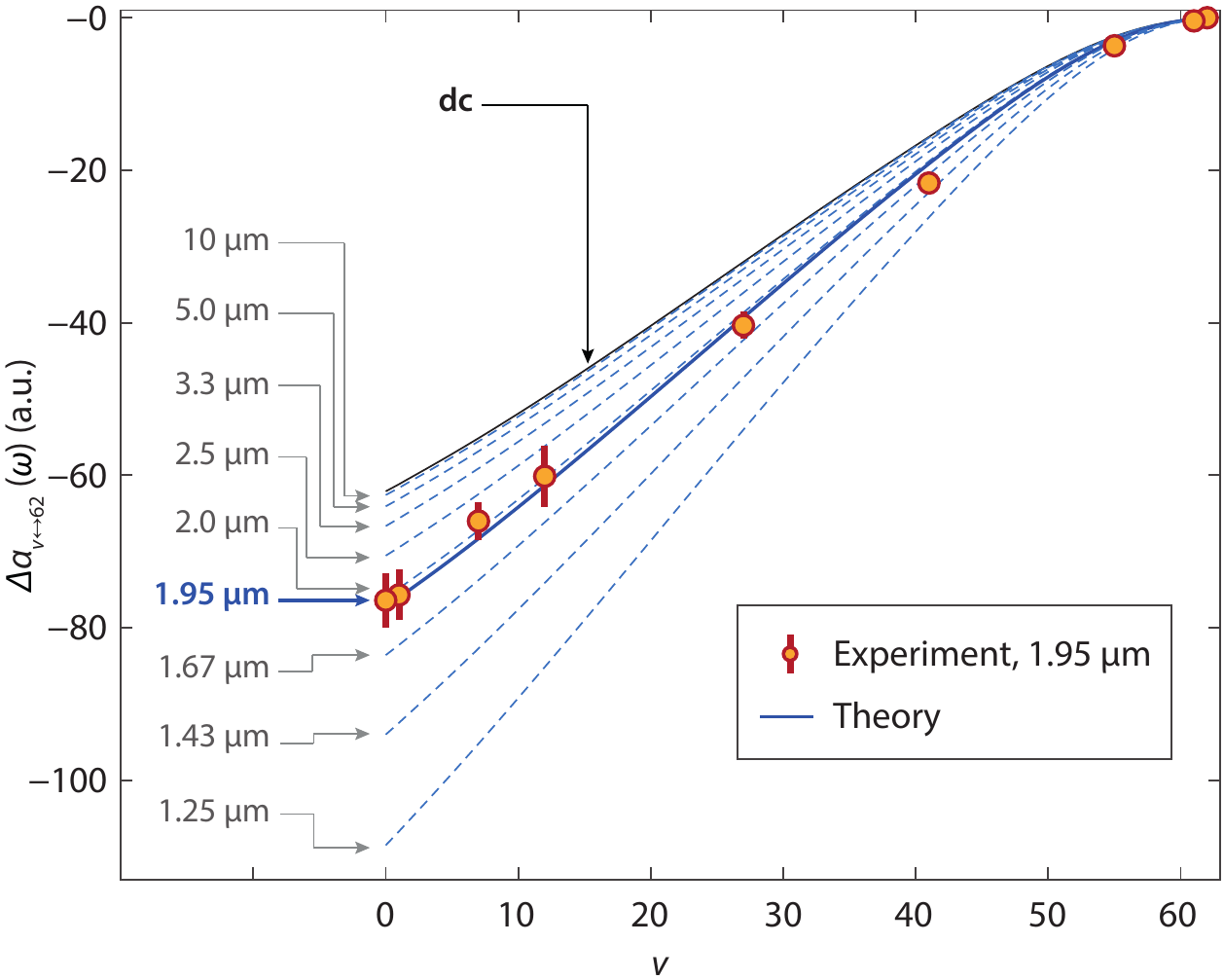}
\caption{Differential polarizability with respect to the least-bound $v=62$ state in ground state Sr$_2$. Points denote experimentally measured ac polarizabilities at $\lambda = 1.95\,$\textmu{}m. Lines are \emph{ab initio} polarizabilities from dc to $\lambda = 1.25\,$\textmu{}m.}
\label{fig:polarizability_vs_v}
\end{figure}

To determine differential polarizabilities we induce ac Stark shifts on these clock transitions using an additional $1.95\,$\textmu{}m laser. We typically observe ac Stark shifts of up to 20~kHz [as shown for $0\leftrightarrow 62$ in Fig.~\ref{fig:cartoon}(b)]. We measure ac Stark shifts of each transition as a function of $1.95\,$\textmu{}m laser power  relative to the $27\leftrightarrow62$ transition [Fig.~\ref{fig:cartoon}(c)]. We do not observe any significant hyperpolarizability~\cite{Leung2023} and therefore we fit a simple proportion. To determine the differential polarizability, we need to adequately characterize the intensity of the $1.95$-\textmu{}m{} light at our molecules. To do so, we compare the ac Stark shift of the $27\leftrightarrow62$ transition to that of the $\Delta{}m=0$ component of atomic intercombination $^1$S$_0$$\rightarrow$$^3$P$_1$ transition with a differential polarizability of $+326.2(3.6)$~a.u. \cite{SafronovaPrivateCommunication}. For our maximum power of 1.7~W, we have an intensity of $6.8$~kW/cm$^2$. For most transitions, this scheme allows us to determine the differential polarizabilities to 5$\%$ as listed in Table~\ref{tab:results} and shown in Fig.~\ref{fig:polarizability_vs_v}. Any thermal shifts stemming from our $5$-$\mu{\rm K}$ sample~\cite{McDonald2015} are negligible (Supplemental Material).

\begin{figure}[b]
\centering
\includegraphics[width=0.95\columnwidth, clip]{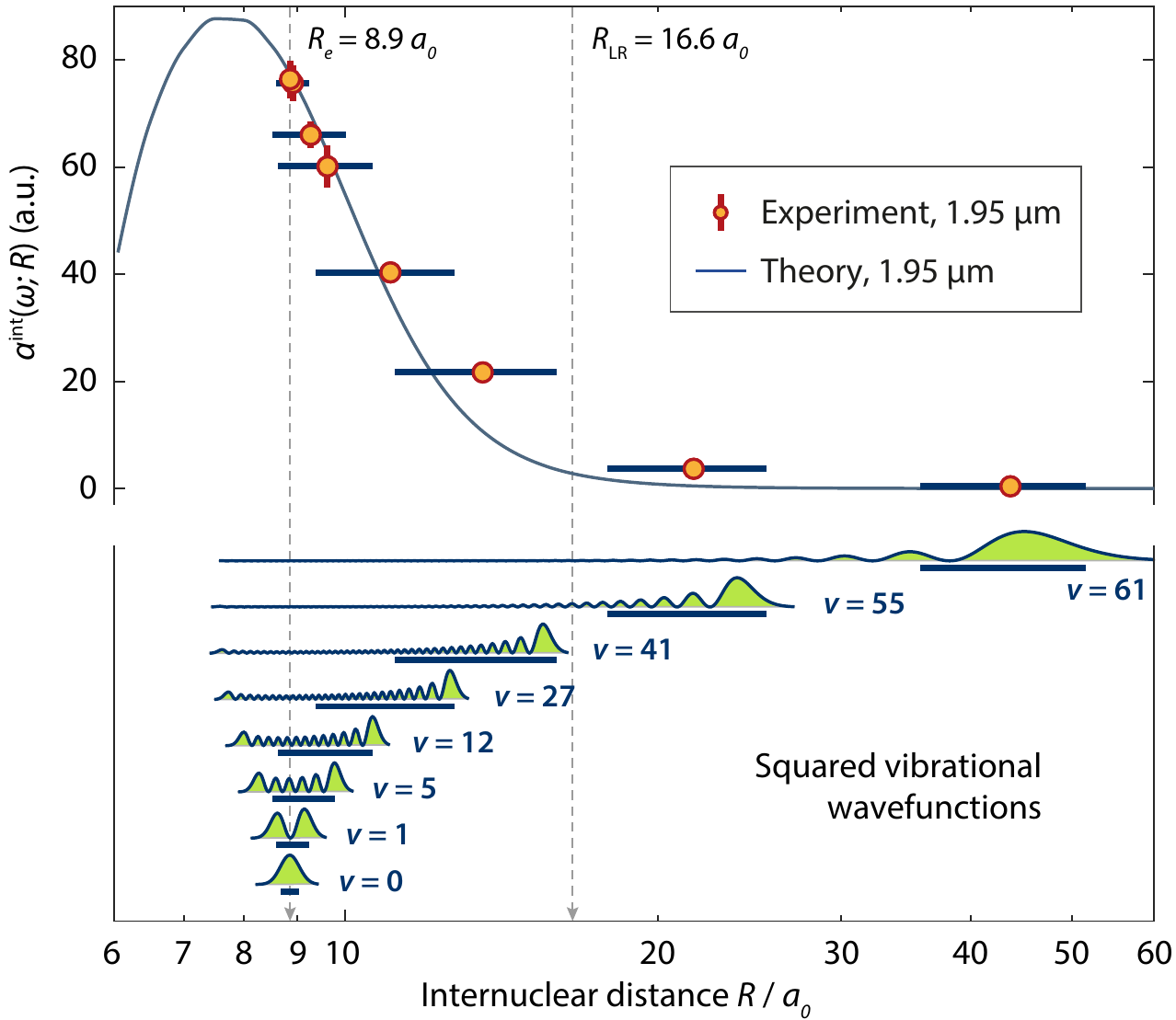}
\caption{Interaction-induced ac polarizability at $\lambda = 1.95$\,\textmu{}m{}. In addition to the \emph{ab initio} result we show absolute experimental polarizabilities in relation to mean internuclear distances $\tilde R$ (Table~\ref{tab:results}). Horizontal bars indicate the range $[\tilde R_v - S_{R_v},\, \tilde R_v + S_{R_v}$] of internuclear distances probed by the vibrational wavefunctions shown in the lower panel. Here $\tilde R_v$ and $S_{R_v}$ are the mean and standard deviation internuclear distances for wavefunction squared treated as a probability distribution. $R_e$ and $R_{\rm LR}$ are the equlibrium distance and the LeRoy radius~\cite{Leroy1970, LeRoy1973}.}
\label{fig:polarizability_vs_R}
\end{figure}

To calculate the BBR shifts we need a model of the differential polarizabilities at all wavelengths from dc to infrared. The overwhelming majority of the BBR spectrum falls below 2~\textmu{}m. While we cannot experimentally probe this entire range of wavelengths, we can leverage close agreement between theory and experiment at $1.95\,$\textmu{}m{} and extend theoretical models to provide a full description of the BBR shift. We use modern quantum chemistry methods to calculate the differential polarizabilities for all molecular clock transitions thusly: first, we calculate \emph{ab initio} electronic polarizabilities of the strontium dimer as a function of internuclear distance $R$, and second, we obtain the polarizability for each vibrational level as an average of the electronic polarizability over the vibrational wavefunction. 

In homonuclear molecules only electronic transitions contribute to polarizabilities and BBR shifts. To calculate the electronic polarizability, we employ the approach based on asymmetric analytical derivative of the coupled-cluster energy with single and double excitations (CCSD)~\cite{Nanda2016}, as implemented in the \mbox{Q-Chem 5} package~\cite{QChem5}. We use the ECP28MDF pseudopotential and its dedicated valence basis set~\cite{Lim2006}. 

For any given light frequency $\omega$, we first calculate the molecular interaction-induced polarizability, $\alpha^{\rm int}_{ij}(\omega; R) = \alpha_{ij}(\omega; R) - 2\alpha_{\rm atom}(\omega)$, where $\alpha_{ij}(\omega)$ are tensor components of the total molecular polarizability and $\alpha_{\rm atom}(\omega)$ is the atomic polarizability at~frequency~$\omega$. Since we only use isotropic $J=0$ states, we take the trace polarizability $\alpha^{\rm int}(\omega; R) = [\alpha^{\rm int}_{zz}(\omega; R) + 2\alpha^{\rm int}_{xx}(\omega; R)]/3$~\cite{Dalgarno1971, Brown2003}. We extend the model for large $R$ using a fitted long-range form $\alpha^{\rm int}(\omega; R) \sim A_6(\omega) R^{-6} + A_8(\omega) R^{-8} + A_{10}(\omega) R^{-10}$~\cite{Heijmen1996}. Figure~\ref{fig:polarizability_vs_R} shows the isotropic component $\alpha^{\rm int}(\omega; R)$ at $1.95\,$\textmu{}m{} as a function of $R$. 

Secondly, we calculate the polarizability of each vibrational level $v$ by averaging the electronic polarizability $\alpha^{\rm int}(R)$ over the level's vibrational wavefunction $\Psi_v(R)$: 
\begin{equation}
  \alpha^{\rm int}_v(\omega) = \int_{0}^{\infty} |\Psi_v(R)|^2 \, \alpha^{\rm int}(\omega; R) dR \label{eq:averaging}
\end{equation}
where the differential polarizability for a transition $v \leftrightarrow v'$ is 
\begin{equation}
    \Delta\alpha_{v \leftrightarrow v'}(\omega) = \alpha^{\rm int}_{v'}(\omega)-\alpha^{\rm int}_{v}(\omega).
\end{equation}
We obtain the vibrational wavefunctions by solving the Schr\"odinger equation, $[-(\hbar^2/2\mu)(d^2/dR^2) + V(R)] \Psi_v(R) = E_v \Psi_v(R)$, using a matrix method~\cite{Colbert1992, Tiesinga1998}. We use an empirical molecular potential $V(R)$~\cite{Stein2010}; the reduced mass $\mu$ equals half the mass of a Sr atom. The uncertainties of the potential curve are negligible for our purposes (Supplemental Material). Figure~\ref{fig:polarizability_vs_v} shows calculated differential dc and ac polarizabilities for $v \leftrightarrow 62$ transitions. It is noteworthy that this approach is valid only when the adiabaticity condition is maintained, that is, that the ground-state potential does not cross any of the excited-state potentials if shifted upwards by the energy of the incident photon. In Sr$_2$, this limits the photon wavenumber to about 8000~cm$^{-1}$ ($1.25$\,\textmu{}m). Both our $1.95$-$\mu$m (5128-cm$^{-1}$) laser and room-temperature BBR are well within this margin.

We first validate the \emph{ab initio} model using polarizabilities of the ground-state Sr atom. At dc we find a polarizability of $+197.327$~a.u., in excellent agreement with the state-of-the-art semi-empirical value of $+197.14(20)$~a.u.~\cite{Safronova2013}. Similarly, our ac polarizability of $+207.524$ a.u. at $1.95$~\textmu{}m{} agrees perfectly with the value of $+208.2(1.1)$~a.u.~\cite{SafronovaPrivateCommunication}.

The key test of our model is the direct comparison and strong agreement of measured differential polarizability at $1.95$\,\textmu{m} with the calculated values (Figure~\ref{fig:polarizability_vs_v}). For example, the theoretical differential polarizability for the $0\leftrightarrow62$ clock transition, $\Delta\alpha_{0 \leftrightarrow 62}(\omega) = -77.2$~a.u. compares well to the experimental  $-76.4(3.6)$~a.u. Moving to more weakly bound target states, the differential polarizabilities decrease monotonically. We elucidate this using the $R$-centroid approximation \cite{Fraser1954} and the concept of a LeRoy radius $R_{\rm LR}$~\cite{Leroy1970, LeRoy1973}. Firstly, the $R$-centroid approximation allows us to estimate the interaction-induced polarizability at the mean internuclear distance $\tilde R_v$ of state $v$ using the differential polarizability of a $v \leftrightarrow 62$ transition:
\begin{equation}
     \alpha^{\rm int}(\omega; \tilde R_{v}) \approx  -\Delta\alpha_{v\leftrightarrow62}(\omega),
\end{equation}
where $\tilde R_v = \int_{0}^{\infty} |\Psi_v(R)|^2 \, R dR$. We neglect the small interaction-induced polarizability of the $v'=62$ state. Thus, different vibrational transitions effectively serve as probes of polarizabilities, each at a different internuclear separation (Figure~\ref{fig:polarizability_vs_R}).

The range of investigated target levels from the ground $v=0$ to the second-to-least bound $v=61$ states spans internuclear distances from $8.86\,a_0$ (approximately the equilibrium distance $R_e$) to $43.6\,a_0$. Beyond the LeRoy radius $R_{\rm LR} = 16.6\,a_0$ the interaction-induced polarizability is negligible: Sr$_2$ becomes a ``physicist's molecule''~\cite{Jones2006} whose polarizability is that of two strontium atoms. At shorter internuclear separations, it becomes a ``chemist's molecule'' and picks up over 80~a.u. of extra polarizability due to molecular bonding of the two consituent atoms. The qualitative boundary between the two extremes is set by $R_{LR}~=~2(r_A+r_B)$ where $r_A=r_B=4.15\,a_0$ are the RMS charge radii of the two atoms \cite{Clementi1967}. By selecting vibrational levels with different mean internuclear distances, we scan the interaction-induced polarizabilities at different internuclear separations, interpolating between the two extremes of ``chemist's'' and ``physicist's'' molecules.

To estimate the relative uncertainty of our theoretical model, we fit it to the experimental data by simple scaling. The best least-squares fit is achieved by scaling the model up by $+1.8(2.4)\%$. This is compatible with zero, showing that no model scaling is necessary; in fact, the reduced chi-square $\chi^2/{\rm dof} = 1.78$ for the scaled model (${\rm dof} = 7$) is worse than $\chi^2/{\rm dof} = 1.69$ (${\rm dof} = 8$) for the original unscaled model. Thus, our \emph{ab initio} model for the molecular polarizability contains no free parameters, justifying its use for all photon wavelengths. Out of caution, we combine the $2.4\%$ uncertainty from the scaling factor with an additional $1.8\%$ possible systematic error to obtain a ``Type B'' uncertainty~\cite{Taylor1994} of $2.6\%$. 

\begin{figure}[t]
\centering
\includegraphics[width=0.98\columnwidth]{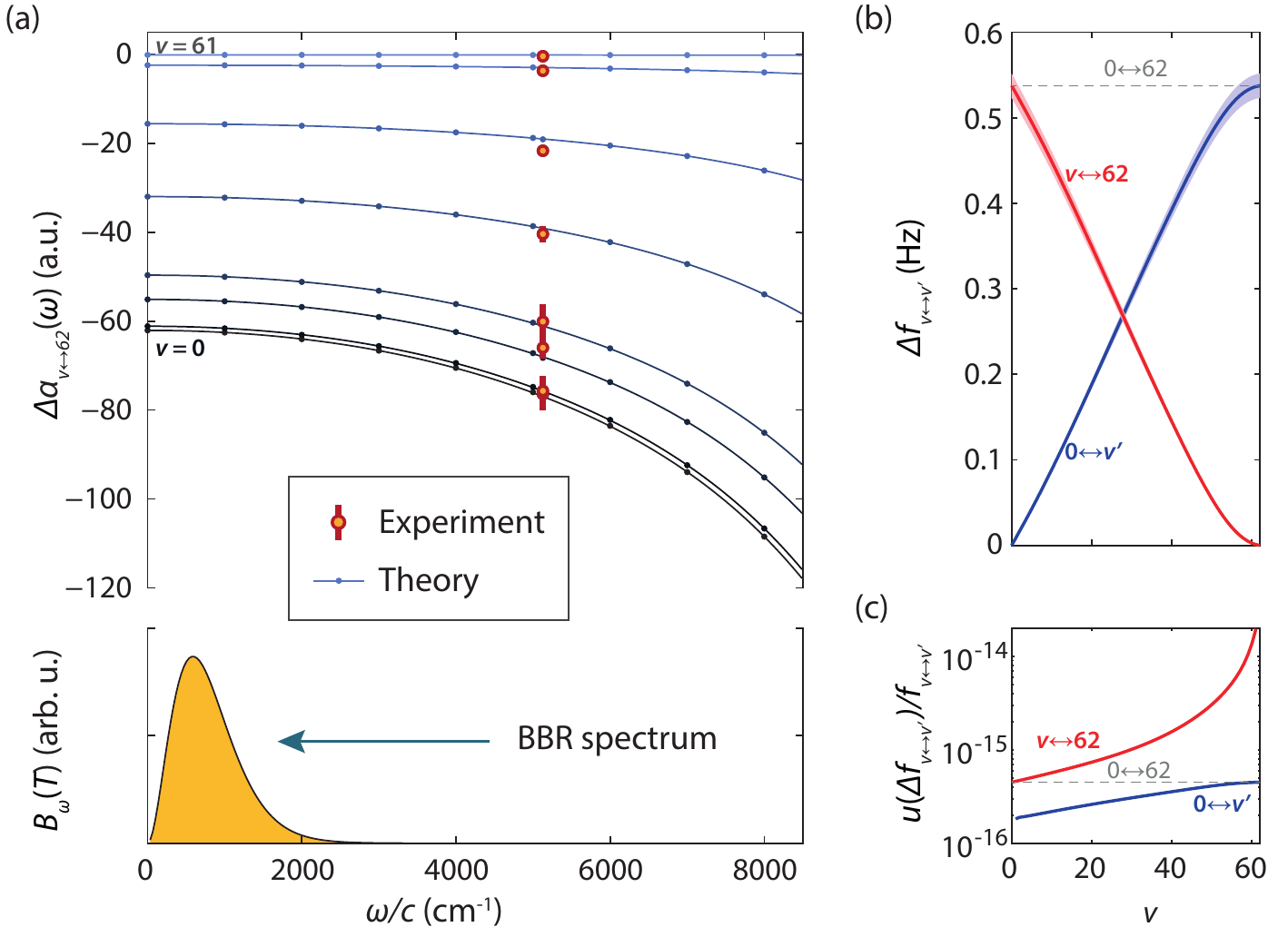}
\caption{(a) Differential polarizabilities in selected clock transitions. Below, a plot of a BBR spectral radiance $B_\omega(T)$ at 300~K. (b) Absolute BBR shift for $0 \leftrightarrow v'$ clock transitions. (c) Relative BBR uncertainty in same clock configurations. \label{fig:BBR} } 
\end{figure}

Finally, we calculate the BBR shift $\Delta f_{v \leftrightarrow v'}$ by integrating the ac Stark shift over the BBR spectrum~\cite{Farley1981, Porsev2006, Middelmann2012}:
\begin{equation}
    \Delta f_{v \leftrightarrow v'} = -\frac{1}{2h}\int_0^\infty\frac{4\pi}{\epsilon_0 c} B_\omega(T) \Delta\alpha_{v\leftrightarrow v'}(\omega)d\omega.
\end{equation}
The BBR spectral radiance at~temperature~$T$ is
\begin{equation}
    B_\omega(T) = \frac{\hbar \omega^3}{4\pi^3 c^2} \frac{1}{\exp(\hbar \omega/k_B T) - 1}.
\end{equation}
Typically, BBR shifts for atomic clocks are determined using sum-over-states to calculate the static and dynamic terms~\cite{Farley1981, Mitroy2010, Safronova2013, Middelmann2012, Lisdat2021}, but we already have computed the dynamic polarizabilities. We can directly integrate the BBR shift. Since practically all of the BBR spectrum falls below any resonance frequencies in our system, we expand the polarizability using Cauchy coefficients~\cite{Mitroy2010}: $\Delta\alpha_{v \leftrightarrow v'}(\omega) = \Delta\alpha_{v \leftrightarrow v'}^{(0)} + \Delta\alpha_{v \leftrightarrow v'}^{(2)}\omega^2 + \Delta\alpha_{v \leftrightarrow v'}^{(4)}\omega^4 + \ldots$ that we fit to tenth order to numerical polarizabilities [Fig.~\ref{fig:BBR}(a)]. This allows expressing the BBR shift as a series:
\begin{equation}
    \Delta f_{v\leftrightarrow v'} = \sum_{n=0,2,\ldots} \Delta f_v^{(n)} =
        \sum_{n=0,2,\ldots} - \frac{c_n \Delta\alpha_{v \leftrightarrow v'}^{(n)}}{4 \pi^3 \epsilon_0 c^3}
          \left(\frac{k_B T}{\hbar} \right)^{4+n},
\end{equation}
where the Planck integrals $c_n = \int_0^\infty u^{3+n}/(e^u -1)du$ appear in Table~\ref{tab:bbr} (Supplemental Material). The leading term is the well known static contribution~\cite{Porsev2006, Safronova2013}, while further terms constitute a dynamic correction $\eta$ on the order of $0.5$--$0.6\,\%$ (Table~\ref{tab:bbr}). Here, terms beyond the second order are negligible.

\begin{table}[t]
    \caption{Contributions to the BBR shift at 300 K for the $0 \leftrightarrow 1$ and $0 \leftrightarrow 62$ transitions. \label{tab:bbr}}
    \begin{ruledtabular}
        \begin{tabular}{cccccc}
            $n$ & $c_n$  & $\Delta f_{0\leftrightarrow1}^{(n)}$ (Hz) & $\Delta f_{0\leftrightarrow1}^{(n)}/f_{0\leftrightarrow1}$ &$\Delta f_{0\leftrightarrow62}^{(n)}$ (Hz) &  $\Delta f_{0\leftrightarrow62}^{(n)}/f_{0\leftrightarrow62}$ \\
            \hline
            0 & $\pi^4/15$ & $+0.0081$ & $+6.8\times10^{-15}$ & $+0.53$ &  $+1.7\times10^{-14}$ \\
            2 & $8\pi^6/63$ & $+6.1\times10^{-5}$ & $+5.1\times10^{-17}$ & $+0.0033$  & $+1.0\times10^{-16}$\\
            4 & $8\pi^8/15$ & $+6.5\times10^{-7}$ & $+5.5\times10^{-19}$ & $+6.3\times10^{-5}$  & $+2.0\times10^{-18}$\\
            
            \hline
            $\eta\,(\%)$ & & \multicolumn{2}{c}{$0.54$} & \multicolumn{2}{c}{$0.62$}
        \end{tabular}
    \end{ruledtabular}
\end{table}

Since the molecular clock uniquely provides an array of available clock states, we calculate the BBR shift for other clock transitions. In Fig.~\ref{fig:BBR}(b), we plot the BBR shift for $v\leftrightarrow 62$ transitions, $\Delta f_{v \leftrightarrow 62}$ (red line). For our previously measured clock transition \cite{Leung2023} $\Delta f_{0 \leftrightarrow 62}= +538(15)$~mHz, giving a BBR contribution to fractional uncertainty of $u(\Delta f_{v\leftrightarrow v'})/ f_{v\leftrightarrow v'} = 4.7 \times 10^{-16}$. Further, the BBR contribution to fractional uncertainty of the molecular clock transition can be reduced by handpicking $0 \leftrightarrow v'$ clock transitions (blue line) between deeply bound vibrational states [Fig.~\ref{fig:BBR}(c)]. This configuration could allow fractional uncertainties as low as 1.8$\times10^{-16}$, a factor of $\sim$2.5 lower than the $0\leftrightarrow 62$ transition. 

Clock transitions between deeply bound states could allow magic wavelengths further detuned from excited molecular resonances due to a smaller polarizability gap to overcome. That could improve molecular trap lifetimes and Q-factors. These can also be improved by switching to vertical lattice geometry. Conversely, this requires the use of STIRAP~\cite{Bergmann1998, Vitanov2017, Leung2021} to initialize the molecule population in a deeply bound state, increasing experimental complexity. 

In the future, our technique can be pushed further. The polarizability measurement relies on frequency shifts that could be determined at the full Hz-level clock accuracy. It also depends on semiempirical atomic polarizabilities that currently contribute about 10\% of the error bar. However, with better measurements, the \textit{ab initio} model will cease to agree with experiment. Scaling is an option, but a complementary approach is possible where polarizabilities are measured at different wavelengths and Cauchy coefficients are instead fitted to experiment.

In conclusion, we have determined the BBR shift in a strontium molecular lattice clock. We leveraged agreement between precision spectroscopy and modern quantum chemistry to provide a robust description of the polarizabilities of ground state Sr$_2$ molecules. Specifically, we performed ac Stark shift spectroscopy of several molecular clock transitions throughout the ground state potential induced by an additional mid-infrared laser. These measurements were in excellent agreement with \textit{ab initio} calculations of molecular polarizability, lending credence to extending this model to other wavelengths. This determination will allow us to control the BBR shift systematic to the 10$^{-16}$ level. Selecting a clock transition between deeply bound vibrational states ($v<10$) could further suppress the BBR effect. Additional measurements of ac or dc Stark shifts, such as by direct application of an electric field \cite{Middelmann2012} or with a CO$_2$ laser \cite{JingbiaoChen2006,Arnold2018}, could further constrain the theoretical model and improve control of the BBR systematic. A next-generation molecular clock could search for new interactions beyond the Standard Model or probe the variations of fundamental constants. This work paves the way towards mHz-level spectroscopy in Sr$_2$ molecules. 

\begin{acknowledgments}
    We thank M. Safronova for providing theoretical atomic polarizabilities, P. S. Żuchowski for fruitful discussions, I. Majewska for involvement and discussions at the early stages of this project and J. Dai, D. Mitra and Q. Sun for experimental assistance. This work was supported by NSF grant PHY-1911959, AFOSR MURI FA9550-21-1-0069, ONR grant N00014-21-1-2644, the Brown Science Foundation, and the Polish National Science Centre (NCN) grant 2016/21/B/ST4/03877.  M. B. was partially funded by the Polish National Agency for Academic Exchange within the Bekker Programme, project PPN/BEK/2020/1/00306/U/00001, and by NCN, grant 2017/25/B/ST4/01486. W.S. acknowledges Polish high-performance computing infrastructure PLGrid (HPC Centers: ACK Cyfronet AGH) for providing computer facilities and support within computational grant no. PLG/2022/015675.
\end{acknowledgments}
\bibliography{library}

\begin{thebibliography}{75}%
\makeatletter
\providecommand \@ifxundefined [1]{%
 \@ifx{#1\undefined}
}%
\providecommand \@ifnum [1]{%
 \ifnum #1\expandafter \@firstoftwo
 \else \expandafter \@secondoftwo
 \fi
}%
\providecommand \@ifx [1]{%
 \ifx #1\expandafter \@firstoftwo
 \else \expandafter \@secondoftwo
 \fi
}%
\providecommand \natexlab [1]{#1}%
\providecommand \enquote  [1]{``#1''}%
\providecommand \bibnamefont  [1]{#1}%
\providecommand \bibfnamefont [1]{#1}%
\providecommand \citenamefont [1]{#1}%
\providecommand \href@noop [0]{\@secondoftwo}%
\providecommand \href [0]{\begingroup \@sanitize@url \@href}%
\providecommand \@href[1]{\@@startlink{#1}\@@href}%
\providecommand \@@href[1]{\endgroup#1\@@endlink}%
\providecommand \@sanitize@url [0]{\catcode `\\12\catcode `\$12\catcode
  `\&12\catcode `\#12\catcode `\^12\catcode `\_12\catcode `\%12\relax}%
\providecommand \@@startlink[1]{}%
\providecommand \@@endlink[0]{}%
\providecommand \url  [0]{\begingroup\@sanitize@url \@url }%
\providecommand \@url [1]{\endgroup\@href {#1}{\urlprefix }}%
\providecommand \urlprefix  [0]{URL }%
\providecommand \Eprint [0]{\href }%
\providecommand \doibase [0]{https://doi.org/}%
\providecommand \selectlanguage [0]{\@gobble}%
\providecommand \bibinfo  [0]{\@secondoftwo}%
\providecommand \bibfield  [0]{\@secondoftwo}%
\providecommand \translation [1]{[#1]}%
\providecommand \BibitemOpen [0]{}%
\providecommand \bibitemStop [0]{}%
\providecommand \bibitemNoStop [0]{.\EOS\space}%
\providecommand \EOS [0]{\spacefactor3000\relax}%
\providecommand \BibitemShut  [1]{\csname bibitem#1\endcsname}%
\let\auto@bib@innerbib\@empty
\bibitem [{\citenamefont {Takamoto}\ \emph {et~al.}(2005)\citenamefont
  {Takamoto}, \citenamefont {Hong}, \citenamefont {Higashi},\ and\
  \citenamefont {Katori}}]{Takamoto2005}%
  \BibitemOpen
  \bibfield  {author} {\bibinfo {author} {\bibfnamefont {M.}~\bibnamefont
  {Takamoto}}, \bibinfo {author} {\bibfnamefont {F.-L.}\ \bibnamefont {Hong}},
  \bibinfo {author} {\bibfnamefont {R.}~\bibnamefont {Higashi}},\ and\ \bibinfo
  {author} {\bibfnamefont {H.}~\bibnamefont {Katori}},\ }\bibfield  {title}
  {\bibinfo {title} {{An optical lattice clock}},\ }\href
  {https://doi.org/10.1038/nature03541} {\bibfield  {journal} {\bibinfo
  {journal} {Nature}\ }\textbf {\bibinfo {volume} {435}},\ \bibinfo {pages}
  {321} (\bibinfo {year} {2005})}\BibitemShut {NoStop}%
\bibitem [{\citenamefont {Katori}(2011)}]{Katori2011}%
  \BibitemOpen
  \bibfield  {author} {\bibinfo {author} {\bibfnamefont {H.}~\bibnamefont
  {Katori}},\ }\bibfield  {title} {\bibinfo {title} {{Optical lattice clocks
  and quantum metrology}},\ }\href {https://doi.org/10.1038/nphoton.2011.45}
  {\bibfield  {journal} {\bibinfo  {journal} {Nature Photonics}\ }\textbf
  {\bibinfo {volume} {5}},\ \bibinfo {pages} {203} (\bibinfo {year}
  {2011})}\BibitemShut {NoStop}%
\bibitem [{\citenamefont {McGrew}\ \emph {et~al.}(2018)\citenamefont {McGrew},
  \citenamefont {Zhang}, \citenamefont {Fasano}, \citenamefont
  {Sch{\"{a}}ffer}, \citenamefont {Beloy}, \citenamefont {Nicolodi},
  \citenamefont {Brown}, \citenamefont {Hinkley}, \citenamefont {Milani},
  \citenamefont {Schioppo}, \citenamefont {Yoon},\ and\ \citenamefont
  {Ludlow}}]{McGrew2018}%
  \BibitemOpen
  \bibfield  {author} {\bibinfo {author} {\bibfnamefont {W.~F.}\ \bibnamefont
  {McGrew}}, \bibinfo {author} {\bibfnamefont {X.}~\bibnamefont {Zhang}},
  \bibinfo {author} {\bibfnamefont {R.~J.}\ \bibnamefont {Fasano}}, \bibinfo
  {author} {\bibfnamefont {S.~A.}\ \bibnamefont {Sch{\"{a}}ffer}}, \bibinfo
  {author} {\bibfnamefont {K.}~\bibnamefont {Beloy}}, \bibinfo {author}
  {\bibfnamefont {D.}~\bibnamefont {Nicolodi}}, \bibinfo {author}
  {\bibfnamefont {R.~C.}\ \bibnamefont {Brown}}, \bibinfo {author}
  {\bibfnamefont {N.}~\bibnamefont {Hinkley}}, \bibinfo {author} {\bibfnamefont
  {G.}~\bibnamefont {Milani}}, \bibinfo {author} {\bibfnamefont
  {M.}~\bibnamefont {Schioppo}}, \bibinfo {author} {\bibfnamefont {T.~H.}\
  \bibnamefont {Yoon}},\ and\ \bibinfo {author} {\bibfnamefont {A.~D.}\
  \bibnamefont {Ludlow}},\ }\bibfield  {title} {\bibinfo {title} {{Atomic clock
  performance enabling geodesy below the centimetre level}},\ }\href
  {https://doi.org/10.1038/s41586-018-0738-2} {\bibfield  {journal} {\bibinfo
  {journal} {Nature}\ }\textbf {\bibinfo {volume} {564}},\ \bibinfo {pages}
  {87} (\bibinfo {year} {2018})}\BibitemShut {NoStop}%
\bibitem [{\citenamefont {Bothwell}\ \emph {et~al.}(2019)\citenamefont
  {Bothwell}, \citenamefont {Kedar}, \citenamefont {Oelker}, \citenamefont
  {Robinson}, \citenamefont {Bromley}, \citenamefont {Tew}, \citenamefont
  {Ye},\ and\ \citenamefont {Kennedy}}]{Bothwell2019}%
  \BibitemOpen
  \bibfield  {author} {\bibinfo {author} {\bibfnamefont {T.}~\bibnamefont
  {Bothwell}}, \bibinfo {author} {\bibfnamefont {D.}~\bibnamefont {Kedar}},
  \bibinfo {author} {\bibfnamefont {E.}~\bibnamefont {Oelker}}, \bibinfo
  {author} {\bibfnamefont {J.~M.}\ \bibnamefont {Robinson}}, \bibinfo {author}
  {\bibfnamefont {S.~L.}\ \bibnamefont {Bromley}}, \bibinfo {author}
  {\bibfnamefont {W.~L.}\ \bibnamefont {Tew}}, \bibinfo {author} {\bibfnamefont
  {J.}~\bibnamefont {Ye}},\ and\ \bibinfo {author} {\bibfnamefont {C.~J.}\
  \bibnamefont {Kennedy}},\ }\bibfield  {title} {\bibinfo {title} {{JILA SrI
  optical lattice clock with uncertainty of $2\times10^{-18}$}},\ }\href
  {https://doi.org/10.1088/1681-7575/ab4089} {\bibfield  {journal} {\bibinfo
  {journal} {Metrologia}\ }\textbf {\bibinfo {volume} {56}},\ \bibinfo {pages}
  {065004} (\bibinfo {year} {2019})}\BibitemShut {NoStop}%
\bibitem [{\citenamefont {McGrew}\ \emph {et~al.}(2019)\citenamefont {McGrew},
  \citenamefont {Zhang}, \citenamefont {Leopardi}, \citenamefont {Fasano},
  \citenamefont {Nicolodi}, \citenamefont {Beloy}, \citenamefont {Yao},
  \citenamefont {Sherman}, \citenamefont {Sch\"{a}ffer}, \citenamefont
  {Savory}, \citenamefont {Brown}, \citenamefont {R\"{o}misch}, \citenamefont
  {Oates}, \citenamefont {Parker}, \citenamefont {Fortier},\ and\ \citenamefont
  {Ludlow}}]{McGrew2019}%
  \BibitemOpen
  \bibfield  {author} {\bibinfo {author} {\bibfnamefont {W.~F.}\ \bibnamefont
  {McGrew}}, \bibinfo {author} {\bibfnamefont {X.}~\bibnamefont {Zhang}},
  \bibinfo {author} {\bibfnamefont {H.}~\bibnamefont {Leopardi}}, \bibinfo
  {author} {\bibfnamefont {R.~J.}\ \bibnamefont {Fasano}}, \bibinfo {author}
  {\bibfnamefont {D.}~\bibnamefont {Nicolodi}}, \bibinfo {author}
  {\bibfnamefont {K.}~\bibnamefont {Beloy}}, \bibinfo {author} {\bibfnamefont
  {J.}~\bibnamefont {Yao}}, \bibinfo {author} {\bibfnamefont {J.~A.}\
  \bibnamefont {Sherman}}, \bibinfo {author} {\bibfnamefont {S.~A.}\
  \bibnamefont {Sch\"{a}ffer}}, \bibinfo {author} {\bibfnamefont
  {J.}~\bibnamefont {Savory}}, \bibinfo {author} {\bibfnamefont {R.~C.}\
  \bibnamefont {Brown}}, \bibinfo {author} {\bibfnamefont {S.}~\bibnamefont
  {R\"{o}misch}}, \bibinfo {author} {\bibfnamefont {C.~W.}\ \bibnamefont
  {Oates}}, \bibinfo {author} {\bibfnamefont {T.~E.}\ \bibnamefont {Parker}},
  \bibinfo {author} {\bibfnamefont {T.~M.}\ \bibnamefont {Fortier}},\ and\
  \bibinfo {author} {\bibfnamefont {A.~D.}\ \bibnamefont {Ludlow}},\ }\bibfield
   {title} {\bibinfo {title} {Towards the optical second: verifying optical
  clocks at the {SI} limit},\ }\href {https://doi.org/10.1364/OPTICA.6.000448}
  {\bibfield  {journal} {\bibinfo  {journal} {Optica}\ }\textbf {\bibinfo
  {volume} {6}},\ \bibinfo {pages} {448} (\bibinfo {year} {2019})}\BibitemShut
  {NoStop}%
\bibitem [{\citenamefont {Lodewyck}(2019)}]{Lodewyck_2019}%
  \BibitemOpen
  \bibfield  {author} {\bibinfo {author} {\bibfnamefont {J.}~\bibnamefont
  {Lodewyck}},\ }\bibfield  {title} {\bibinfo {title} {On a definition of the
  si second with a set of optical clock transitions},\ }\href
  {https://doi.org/10.1088/1681-7575/ab3a82} {\bibfield  {journal} {\bibinfo
  {journal} {Metrologia}\ }\textbf {\bibinfo {volume} {56}},\ \bibinfo {pages}
  {055009} (\bibinfo {year} {2019})}\BibitemShut {NoStop}%
\bibitem [{\citenamefont {Bize}(2019)}]{BIZE2019153}%
  \BibitemOpen
  \bibfield  {author} {\bibinfo {author} {\bibfnamefont {S.}~\bibnamefont
  {Bize}},\ }\bibfield  {title} {\bibinfo {title} {The unit of time: Present
  and future directions},\ }\href
  {https://doi.org/https://doi.org/10.1016/j.crhy.2019.02.002} {\bibfield
  {journal} {\bibinfo  {journal} {Comptes Rendus Physique}\ }\textbf {\bibinfo
  {volume} {20}},\ \bibinfo {pages} {153} (\bibinfo {year} {2019})}\BibitemShut
  {NoStop}%
\bibitem [{\citenamefont {Borkowski}(2018)}]{Borkowski2018}%
  \BibitemOpen
  \bibfield  {author} {\bibinfo {author} {\bibfnamefont {M.}~\bibnamefont
  {Borkowski}},\ }\bibfield  {title} {\bibinfo {title} {{Optical Lattice Clocks
  with Weakly Bound Molecules}},\ }\href
  {https://doi.org/10.1103/PhysRevLett.120.083202} {\bibfield  {journal}
  {\bibinfo  {journal} {Physical Review Letters}\ }\textbf {\bibinfo {volume}
  {120}},\ \bibinfo {pages} {083202} (\bibinfo {year} {2018})}\BibitemShut
  {NoStop}%
\bibitem [{\citenamefont {Kondov}\ \emph {et~al.}(2019)\citenamefont {Kondov},
  \citenamefont {Lee}, \citenamefont {Leung}, \citenamefont {Liedl},
  \citenamefont {Majewska}, \citenamefont {Moszynski},\ and\ \citenamefont
  {Zelevinsky}}]{Kondov2019a}%
  \BibitemOpen
  \bibfield  {author} {\bibinfo {author} {\bibfnamefont {S.~S.}\ \bibnamefont
  {Kondov}}, \bibinfo {author} {\bibfnamefont {C.-H.}\ \bibnamefont {Lee}},
  \bibinfo {author} {\bibfnamefont {K.~H.}\ \bibnamefont {Leung}}, \bibinfo
  {author} {\bibfnamefont {C.}~\bibnamefont {Liedl}}, \bibinfo {author}
  {\bibfnamefont {I.}~\bibnamefont {Majewska}}, \bibinfo {author}
  {\bibfnamefont {R.}~\bibnamefont {Moszynski}},\ and\ \bibinfo {author}
  {\bibfnamefont {T.}~\bibnamefont {Zelevinsky}},\ }\bibfield  {title}
  {\bibinfo {title} {Molecular lattice clock with long vibrational coherence},\
  }\href {https://doi.org/10.1038/s41567-019-0632-3} {\bibfield  {journal}
  {\bibinfo  {journal} {Nature Physics}\ }\textbf {\bibinfo {volume} {15}},\
  \bibinfo {pages} {1118} (\bibinfo {year} {2019})}\BibitemShut {NoStop}%
\bibitem [{\citenamefont {Kobayashi}\ \emph {et~al.}(2019)\citenamefont
  {Kobayashi}, \citenamefont {Ogino},\ and\ \citenamefont
  {Inouye}}]{Kobayashi2019a}%
  \BibitemOpen
  \bibfield  {author} {\bibinfo {author} {\bibfnamefont {J.}~\bibnamefont
  {Kobayashi}}, \bibinfo {author} {\bibfnamefont {A.}~\bibnamefont {Ogino}},\
  and\ \bibinfo {author} {\bibfnamefont {S.}~\bibnamefont {Inouye}},\
  }\bibfield  {title} {\bibinfo {title} {{Measurement of the variation of
  electron-to-proton mass ratio using ultracold molecules produced from
  laser-cooled atoms}},\ }\href {https://doi.org/10.1038/s41467-019-11761-1}
  {\bibfield  {journal} {\bibinfo  {journal} {Nature Communications}\ }\textbf
  {\bibinfo {volume} {10}},\ \bibinfo {pages} {3771} (\bibinfo {year}
  {2019})}\BibitemShut {NoStop}%
\bibitem [{\citenamefont {Hanneke}\ \emph {et~al.}(2020)\citenamefont
  {Hanneke}, \citenamefont {Kuzhan},\ and\ \citenamefont
  {Lunstad}}]{Hanneke2021}%
  \BibitemOpen
  \bibfield  {author} {\bibinfo {author} {\bibfnamefont {D.}~\bibnamefont
  {Hanneke}}, \bibinfo {author} {\bibfnamefont {B.}~\bibnamefont {Kuzhan}},\
  and\ \bibinfo {author} {\bibfnamefont {A.}~\bibnamefont {Lunstad}},\
  }\bibfield  {title} {\bibinfo {title} {Optical clocks based on molecular
  vibrations as probes of variation of the proton-to-electron mass ratio},\
  }\href {https://doi.org/10.1088/2058-9565/abc863} {\bibfield  {journal}
  {\bibinfo  {journal} {Quantum Science and Technology}\ }\textbf {\bibinfo
  {volume} {6}},\ \bibinfo {pages} {014005} (\bibinfo {year}
  {2020})}\BibitemShut {NoStop}%
\bibitem [{\citenamefont {Barontini}\ \emph {et~al.}(2022)\citenamefont
  {Barontini}, \citenamefont {Blackburn}, \citenamefont {Boyer}, \citenamefont
  {Butuc-Mayer}, \citenamefont {Calmet}, \citenamefont {Crespo
  L{\'o}pez-Urrutia}, \citenamefont {Curtis}, \citenamefont {Darqui{\'e}},
  \citenamefont {Dunningham}, \citenamefont {Fitch}, \citenamefont {Forgan},
  \citenamefont {Georgiou}, \citenamefont {Gill}, \citenamefont {Godun},
  \citenamefont {Goldwin} \emph {et~al.}}]{Barontini2022}%
  \BibitemOpen
  \bibfield  {author} {\bibinfo {author} {\bibfnamefont {G.}~\bibnamefont
  {Barontini}}, \bibinfo {author} {\bibfnamefont {L.}~\bibnamefont
  {Blackburn}}, \bibinfo {author} {\bibfnamefont {V.}~\bibnamefont {Boyer}},
  \bibinfo {author} {\bibfnamefont {F.}~\bibnamefont {Butuc-Mayer}}, \bibinfo
  {author} {\bibfnamefont {X.}~\bibnamefont {Calmet}}, \bibinfo {author}
  {\bibfnamefont {J.~R.}\ \bibnamefont {Crespo L{\'o}pez-Urrutia}}, \bibinfo
  {author} {\bibfnamefont {E.~A.}\ \bibnamefont {Curtis}}, \bibinfo {author}
  {\bibfnamefont {B.}~\bibnamefont {Darqui{\'e}}}, \bibinfo {author}
  {\bibfnamefont {J.}~\bibnamefont {Dunningham}}, \bibinfo {author}
  {\bibfnamefont {N.~J.}\ \bibnamefont {Fitch}}, \bibinfo {author}
  {\bibfnamefont {E.~M.}\ \bibnamefont {Forgan}}, \bibinfo {author}
  {\bibfnamefont {K.}~\bibnamefont {Georgiou}}, \bibinfo {author}
  {\bibfnamefont {P.}~\bibnamefont {Gill}}, \bibinfo {author} {\bibfnamefont
  {R.~M.}\ \bibnamefont {Godun}}, \bibinfo {author} {\bibfnamefont
  {J.}~\bibnamefont {Goldwin}}, \emph {et~al.},\ }\bibfield  {title} {\bibinfo
  {title} {Measuring the stability of fundamental constants with a network of
  clocks},\ }\href
  {https://epjquantumtechnology.springeropen.com/articles/10.1140/epjqt/s40507-022-00130-5}
  {\bibfield  {journal} {\bibinfo  {journal} {EPJ Quantum Technology}\ }\textbf
  {\bibinfo {volume} {9}},\ \bibinfo {pages} {12} (\bibinfo {year}
  {2022})}\BibitemShut {NoStop}%
\bibitem [{\citenamefont {Salumbides}\ \emph {et~al.}(2014)\citenamefont
  {Salumbides}, \citenamefont {Ubachs},\ and\ \citenamefont
  {Korobov}}]{Salumbides2014}%
  \BibitemOpen
  \bibfield  {author} {\bibinfo {author} {\bibfnamefont {E.~J.}\ \bibnamefont
  {Salumbides}}, \bibinfo {author} {\bibfnamefont {W.}~\bibnamefont {Ubachs}},\
  and\ \bibinfo {author} {\bibfnamefont {V.~I.}\ \bibnamefont {Korobov}},\
  }\bibfield  {title} {\bibinfo {title} {{{Bounds on fifth forces at the
  sub-\AA~length scale}}},\ }\href {https://doi.org/10.1016/j.jms.2014.04.003}
  {\bibfield  {journal} {\bibinfo  {journal} {Journal of Molecular
  Spectroscopy}\ }\textbf {\bibinfo {volume} {300}},\ \bibinfo {pages} {65}
  (\bibinfo {year} {2014})}\BibitemShut {NoStop}%
\bibitem [{\citenamefont {Biesheuvel}\ \emph {et~al.}(2016)\citenamefont
  {Biesheuvel}, \citenamefont {Karr}, \citenamefont {Hilico}, \citenamefont
  {Eikema}, \citenamefont {Ubachs},\ and\ \citenamefont
  {Koelemeij}}]{biesheuvel2016probing}%
  \BibitemOpen
  \bibfield  {author} {\bibinfo {author} {\bibfnamefont {J.}~\bibnamefont
  {Biesheuvel}}, \bibinfo {author} {\bibfnamefont {J.-P.}\ \bibnamefont
  {Karr}}, \bibinfo {author} {\bibfnamefont {L.}~\bibnamefont {Hilico}},
  \bibinfo {author} {\bibfnamefont {K.}~\bibnamefont {Eikema}}, \bibinfo
  {author} {\bibfnamefont {W.}~\bibnamefont {Ubachs}},\ and\ \bibinfo {author}
  {\bibfnamefont {J.}~\bibnamefont {Koelemeij}},\ }\bibfield  {title} {\bibinfo
  {title} {{Probing QED and fundamental constants through laser spectroscopy of
  vibrational transitions in HD$^+$}},\ }\href
  {https://doi.org/10.1038/ncomms10385} {\bibfield  {journal} {\bibinfo
  {journal} {{Nature Communications}}\ }\textbf {\bibinfo {volume} {7}},\
  \bibinfo {pages} {10385} (\bibinfo {year} {2016})}\BibitemShut {NoStop}%
\bibitem [{\citenamefont {Borkowski}\ \emph {et~al.}(2019)\citenamefont
  {Borkowski}, \citenamefont {Buchachenko}, \citenamefont {Ciury{\l}o},
  \citenamefont {Julienne}, \citenamefont {Yamada}, \citenamefont {Kikuchi},
  \citenamefont {Takasu},\ and\ \citenamefont {Takahashi}}]{Borkowski2019}%
  \BibitemOpen
  \bibfield  {author} {\bibinfo {author} {\bibfnamefont {M.}~\bibnamefont
  {Borkowski}}, \bibinfo {author} {\bibfnamefont {A.~A.}\ \bibnamefont
  {Buchachenko}}, \bibinfo {author} {\bibfnamefont {R.}~\bibnamefont
  {Ciury{\l}o}}, \bibinfo {author} {\bibfnamefont {P.~S.}\ \bibnamefont
  {Julienne}}, \bibinfo {author} {\bibfnamefont {H.}~\bibnamefont {Yamada}},
  \bibinfo {author} {\bibfnamefont {Y.}~\bibnamefont {Kikuchi}}, \bibinfo
  {author} {\bibfnamefont {Y.}~\bibnamefont {Takasu}},\ and\ \bibinfo {author}
  {\bibfnamefont {Y.}~\bibnamefont {Takahashi}},\ }\bibfield  {title} {\bibinfo
  {title} {{Weakly bound molecules as sensors of new gravitylike forces}},\
  }\href {https://doi.org/10.1038/s41598-019-51346-y} {\bibfield  {journal}
  {\bibinfo  {journal} {Scientific Reports}\ }\textbf {\bibinfo {volume} {9}},\
  \bibinfo {pages} {14807} (\bibinfo {year} {2019})}\BibitemShut {NoStop}%
\bibitem [{\citenamefont {Heacock}\ \emph {et~al.}(2021)\citenamefont
  {Heacock}, \citenamefont {Fujiie}, \citenamefont {Haun}, \citenamefont
  {Henins}, \citenamefont {Hirota}, \citenamefont {Hosobata}, \citenamefont
  {Huber}, \citenamefont {Kitaguchi}, \citenamefont {Pushin}, \citenamefont
  {Shimizu}, \citenamefont {Takeda}, \citenamefont {Valdillez}, \citenamefont
  {Yamagata},\ and\ \citenamefont {Young}}]{Heacock2021}%
  \BibitemOpen
  \bibfield  {author} {\bibinfo {author} {\bibfnamefont {B.}~\bibnamefont
  {Heacock}}, \bibinfo {author} {\bibfnamefont {T.}~\bibnamefont {Fujiie}},
  \bibinfo {author} {\bibfnamefont {R.~W.}\ \bibnamefont {Haun}}, \bibinfo
  {author} {\bibfnamefont {A.}~\bibnamefont {Henins}}, \bibinfo {author}
  {\bibfnamefont {K.}~\bibnamefont {Hirota}}, \bibinfo {author} {\bibfnamefont
  {T.}~\bibnamefont {Hosobata}}, \bibinfo {author} {\bibfnamefont {M.~G.}\
  \bibnamefont {Huber}}, \bibinfo {author} {\bibfnamefont {M.}~\bibnamefont
  {Kitaguchi}}, \bibinfo {author} {\bibfnamefont {D.~A.}\ \bibnamefont
  {Pushin}}, \bibinfo {author} {\bibfnamefont {H.}~\bibnamefont {Shimizu}},
  \bibinfo {author} {\bibfnamefont {M.}~\bibnamefont {Takeda}}, \bibinfo
  {author} {\bibfnamefont {R.}~\bibnamefont {Valdillez}}, \bibinfo {author}
  {\bibfnamefont {Y.}~\bibnamefont {Yamagata}},\ and\ \bibinfo {author}
  {\bibfnamefont {A.~R.}\ \bibnamefont {Young}},\ }\bibfield  {title} {\bibinfo
  {title} {Pendell\"{o}sung interferometry probes the neutron charge radius,
  lattice dynamics, and fifth forces},\ }\href
  {https://doi.org/10.1126/science.abc2794} {\bibfield  {journal} {\bibinfo
  {journal} {Science}\ }\textbf {\bibinfo {volume} {373}},\ \bibinfo {pages}
  {1239} (\bibinfo {year} {2021})}\BibitemShut {NoStop}%
\bibitem [{\citenamefont {Schiller}\ and\ \citenamefont
  {Korobov}(2005)}]{Schiller2005}%
  \BibitemOpen
  \bibfield  {author} {\bibinfo {author} {\bibfnamefont {S.}~\bibnamefont
  {Schiller}}\ and\ \bibinfo {author} {\bibfnamefont {V.}~\bibnamefont
  {Korobov}},\ }\bibfield  {title} {\bibinfo {title} {Tests of time
  independence of the electron and nuclear masses with ultracold molecules},\
  }\href {https://doi.org/10.1103/PhysRevA.71.032505} {\bibfield  {journal}
  {\bibinfo  {journal} {Phys. Rev. A}\ }\textbf {\bibinfo {volume} {71}},\
  \bibinfo {pages} {032505} (\bibinfo {year} {2005})}\BibitemShut {NoStop}%
\bibitem [{\citenamefont {{DeMille}}\ \emph {et~al.}(2008)\citenamefont
  {{DeMille}}, \citenamefont {Sainis}, \citenamefont {Sage}, \citenamefont
  {Bergeman}, \citenamefont {Kotochigova},\ and\ \citenamefont
  {Tiesinga}}]{Demille2008}%
  \BibitemOpen
  \bibfield  {author} {\bibinfo {author} {\bibfnamefont {D.}~\bibnamefont
  {{DeMille}}}, \bibinfo {author} {\bibfnamefont {S.}~\bibnamefont {Sainis}},
  \bibinfo {author} {\bibfnamefont {J.}~\bibnamefont {Sage}}, \bibinfo {author}
  {\bibfnamefont {T.}~\bibnamefont {Bergeman}}, \bibinfo {author}
  {\bibfnamefont {S.}~\bibnamefont {Kotochigova}},\ and\ \bibinfo {author}
  {\bibfnamefont {E.}~\bibnamefont {Tiesinga}},\ }\bibfield  {title} {\bibinfo
  {title} {{Enhanced sensitivity to variation of me/mp in molecular pectra}},\
  }\href {https://doi.org/10.1103/PhysRevLett.100.043202} {\bibfield  {journal}
  {\bibinfo  {journal} {Physical Review Letters}\ }\textbf {\bibinfo {volume}
  {100}},\ \bibinfo {pages} {043202} (\bibinfo {year} {2008})}\BibitemShut
  {NoStop}%
\bibitem [{\citenamefont {Beloy}\ \emph {et~al.}(2011)\citenamefont {Beloy},
  \citenamefont {Hauser}, \citenamefont {Borschevsky}, \citenamefont
  {Flambaum},\ and\ \citenamefont {Schwerdtfeger}}]{Beloy2011}%
  \BibitemOpen
  \bibfield  {author} {\bibinfo {author} {\bibfnamefont {K.}~\bibnamefont
  {Beloy}}, \bibinfo {author} {\bibfnamefont {A.~W.}\ \bibnamefont {Hauser}},
  \bibinfo {author} {\bibfnamefont {A.}~\bibnamefont {Borschevsky}}, \bibinfo
  {author} {\bibfnamefont {V.~V.}\ \bibnamefont {Flambaum}},\ and\ \bibinfo
  {author} {\bibfnamefont {P.}~\bibnamefont {Schwerdtfeger}},\ }\bibfield
  {title} {\bibinfo {title} {{Effect of Alpha variation on the vibrational
  spectrum of Sr$_2$}},\ }\href {https://doi.org/10.1103/PhysRevA.84.062114}
  {\bibfield  {journal} {\bibinfo  {journal} {Phys. Rev. A}\ }\textbf {\bibinfo
  {volume} {84}},\ \bibinfo {pages} {062114} (\bibinfo {year}
  {2011})}\BibitemShut {NoStop}%
\bibitem [{\citenamefont {Schiller}\ \emph {et~al.}(2014)\citenamefont
  {Schiller}, \citenamefont {Bakalov},\ and\ \citenamefont
  {Korobov}}]{Schiller2014}%
  \BibitemOpen
  \bibfield  {author} {\bibinfo {author} {\bibfnamefont {S.}~\bibnamefont
  {Schiller}}, \bibinfo {author} {\bibfnamefont {D.}~\bibnamefont {Bakalov}},\
  and\ \bibinfo {author} {\bibfnamefont {V.~I.}\ \bibnamefont {Korobov}},\
  }\bibfield  {title} {\bibinfo {title} {Simplest molecules as candidates for
  precise optical clocks},\ }\href
  {https://doi.org/10.1103/PhysRevLett.113.023004} {\bibfield  {journal}
  {\bibinfo  {journal} {Phys. Rev. Lett.}\ }\textbf {\bibinfo {volume} {113}},\
  \bibinfo {pages} {023004} (\bibinfo {year} {2014})}\BibitemShut {NoStop}%
\bibitem [{\citenamefont {Kajita}\ \emph {et~al.}(2014)\citenamefont {Kajita},
  \citenamefont {Gopakumar}, \citenamefont {Abe}, \citenamefont {Hada},\ and\
  \citenamefont {Keller}}]{Kajita2014}%
  \BibitemOpen
  \bibfield  {author} {\bibinfo {author} {\bibfnamefont {M.}~\bibnamefont
  {Kajita}}, \bibinfo {author} {\bibfnamefont {G.}~\bibnamefont {Gopakumar}},
  \bibinfo {author} {\bibfnamefont {M.}~\bibnamefont {Abe}}, \bibinfo {author}
  {\bibfnamefont {M.}~\bibnamefont {Hada}},\ and\ \bibinfo {author}
  {\bibfnamefont {M.}~\bibnamefont {Keller}},\ }\bibfield  {title} {\bibinfo
  {title} {Test of ${m}_{p}$/${m}_{e}$ changes using vibrational transitions in
  n${}_{2}{}^{+}$},\ }\href {https://doi.org/10.1103/PhysRevA.89.032509}
  {\bibfield  {journal} {\bibinfo  {journal} {Phys. Rev. A}\ }\textbf {\bibinfo
  {volume} {89}},\ \bibinfo {pages} {032509} (\bibinfo {year}
  {2014})}\BibitemShut {NoStop}%
\bibitem [{\citenamefont {Germann}\ \emph {et~al.}(2014)\citenamefont
  {Germann}, \citenamefont {Tong},\ and\ \citenamefont
  {Willitsch}}]{Germann2014}%
  \BibitemOpen
  \bibfield  {author} {\bibinfo {author} {\bibfnamefont {M.}~\bibnamefont
  {Germann}}, \bibinfo {author} {\bibfnamefont {X.}~\bibnamefont {Tong}},\ and\
  \bibinfo {author} {\bibfnamefont {S.}~\bibnamefont {Willitsch}},\ }\bibfield
  {title} {\bibinfo {title} {{Observation of electric-dipole-forbidden infrared
  transitions in cold molecular ions}},\ }\href
  {https://doi.org/10.1038/nphys3085} {\bibfield  {journal} {\bibinfo
  {journal} {Nature Physics}\ }\textbf {\bibinfo {volume} {10}},\ \bibinfo
  {pages} {820} (\bibinfo {year} {2014})}\BibitemShut {NoStop}%
\bibitem [{\citenamefont {Wcis{\l}o}\ \emph {et~al.}(2018)\citenamefont
  {Wcis{\l}o}, \citenamefont {Ablewski}, \citenamefont {Beloy}, \citenamefont
  {Bilicki}, \citenamefont {Bober}, \citenamefont {Brown}, \citenamefont
  {Fasano}, \citenamefont {Ciury{\l}o}, \citenamefont {Hachisu}, \citenamefont
  {Ido}, \citenamefont {Lodewyck}, \citenamefont {Ludlow}, \citenamefont
  {McGrew}, \citenamefont {Morzy{\'{n}}ski}, \citenamefont {Nicolodi},
  \citenamefont {Schioppo}, \citenamefont {Sekido}, \citenamefont {{Le
  Targat}}, \citenamefont {Wolf}, \citenamefont {Zhang}, \citenamefont
  {Zjawin},\ and\ \citenamefont {Zawada}}]{Wcislo2018}%
  \BibitemOpen
  \bibfield  {author} {\bibinfo {author} {\bibfnamefont {P.}~\bibnamefont
  {Wcis{\l}o}}, \bibinfo {author} {\bibfnamefont {P.}~\bibnamefont {Ablewski}},
  \bibinfo {author} {\bibfnamefont {K.}~\bibnamefont {Beloy}}, \bibinfo
  {author} {\bibfnamefont {S.}~\bibnamefont {Bilicki}}, \bibinfo {author}
  {\bibfnamefont {M.}~\bibnamefont {Bober}}, \bibinfo {author} {\bibfnamefont
  {R.}~\bibnamefont {Brown}}, \bibinfo {author} {\bibfnamefont
  {R.}~\bibnamefont {Fasano}}, \bibinfo {author} {\bibfnamefont
  {R.}~\bibnamefont {Ciury{\l}o}}, \bibinfo {author} {\bibfnamefont
  {H.}~\bibnamefont {Hachisu}}, \bibinfo {author} {\bibfnamefont
  {T.}~\bibnamefont {Ido}}, \bibinfo {author} {\bibfnamefont {J.}~\bibnamefont
  {Lodewyck}}, \bibinfo {author} {\bibfnamefont {A.}~\bibnamefont {Ludlow}},
  \bibinfo {author} {\bibfnamefont {W.}~\bibnamefont {McGrew}}, \bibinfo
  {author} {\bibfnamefont {P.}~\bibnamefont {Morzy{\'{n}}ski}}, \bibinfo
  {author} {\bibfnamefont {D.}~\bibnamefont {Nicolodi}}, \bibinfo {author}
  {\bibfnamefont {M.}~\bibnamefont {Schioppo}}, \bibinfo {author}
  {\bibfnamefont {M.}~\bibnamefont {Sekido}}, \bibinfo {author} {\bibfnamefont
  {R.}~\bibnamefont {{Le Targat}}}, \bibinfo {author} {\bibfnamefont
  {P.}~\bibnamefont {Wolf}}, \bibinfo {author} {\bibfnamefont {X.}~\bibnamefont
  {Zhang}}, \bibinfo {author} {\bibfnamefont {B.}~\bibnamefont {Zjawin}},\ and\
  \bibinfo {author} {\bibfnamefont {M.}~\bibnamefont {Zawada}},\ }\bibfield
  {title} {\bibinfo {title} {{New bounds on dark matter coupling from a global
  network of optical atomic clocks}},\ }\href
  {https://doi.org/10.1126/sciadv.aau4869} {\bibfield  {journal} {\bibinfo
  {journal} {Science Advances}\ }\textbf {\bibinfo {volume} {4}},\ \bibinfo
  {pages} {eaau4869} (\bibinfo {year} {2018})}\BibitemShut {NoStop}%
\bibitem [{\citenamefont {Safronova}(2019)}]{Safronova2019}%
  \BibitemOpen
  \bibfield  {author} {\bibinfo {author} {\bibfnamefont {M.~S.}\ \bibnamefont
  {Safronova}},\ }\bibfield  {title} {\bibinfo {title} {The search for
  variation of fundamental constants with clocks},\ }\href
  {https://doi.org/https://doi.org/10.1002/andp.201800364} {\bibfield
  {journal} {\bibinfo  {journal} {Annalen der Physik}\ }\textbf {\bibinfo
  {volume} {531}},\ \bibinfo {pages} {1800364} (\bibinfo {year}
  {2019})}\BibitemShut {NoStop}%
\bibitem [{\citenamefont {Hutzler}(2020)}]{Hutzler2020}%
  \BibitemOpen
  \bibfield  {author} {\bibinfo {author} {\bibfnamefont {N.~R.}\ \bibnamefont
  {Hutzler}},\ }\bibfield  {title} {\bibinfo {title} {Polyatomic molecules as
  quantum sensors for fundamental physics},\ }\href
  {https://doi.org/10.1088/2058-9565/abb9c5} {\bibfield  {journal} {\bibinfo
  {journal} {Quantum Science and Technology}\ }\textbf {\bibinfo {volume}
  {5}},\ \bibinfo {pages} {044011} (\bibinfo {year} {2020})}\BibitemShut
  {NoStop}%
\bibitem [{\citenamefont {Lange}\ \emph {et~al.}(2021)\citenamefont {Lange},
  \citenamefont {Huntemann}, \citenamefont {Rahm}, \citenamefont {Sanner},
  \citenamefont {Shao}, \citenamefont {Lipphardt}, \citenamefont {Tamm},
  \citenamefont {Weyers},\ and\ \citenamefont {Peik}}]{Lange2021}%
  \BibitemOpen
  \bibfield  {author} {\bibinfo {author} {\bibfnamefont {R.}~\bibnamefont
  {Lange}}, \bibinfo {author} {\bibfnamefont {N.}~\bibnamefont {Huntemann}},
  \bibinfo {author} {\bibfnamefont {J.~M.}\ \bibnamefont {Rahm}}, \bibinfo
  {author} {\bibfnamefont {C.}~\bibnamefont {Sanner}}, \bibinfo {author}
  {\bibfnamefont {H.}~\bibnamefont {Shao}}, \bibinfo {author} {\bibfnamefont
  {B.}~\bibnamefont {Lipphardt}}, \bibinfo {author} {\bibfnamefont
  {C.}~\bibnamefont {Tamm}}, \bibinfo {author} {\bibfnamefont {S.}~\bibnamefont
  {Weyers}},\ and\ \bibinfo {author} {\bibfnamefont {E.}~\bibnamefont {Peik}},\
  }\bibfield  {title} {\bibinfo {title} {Improved limits for violations of
  local position invariance from atomic clock comparisons},\ }\href
  {https://doi.org/10.1103/PhysRevLett.126.011102} {\bibfield  {journal}
  {\bibinfo  {journal} {Phys. Rev. Lett.}\ }\textbf {\bibinfo {volume} {126}},\
  \bibinfo {pages} {011102} (\bibinfo {year} {2021})}\BibitemShut {NoStop}%
\bibitem [{\citenamefont {{Le Targat}}\ \emph {et~al.}(2013)\citenamefont {{Le
  Targat}}, \citenamefont {Lorini}, \citenamefont {{Le Coq}}, \citenamefont
  {Zawada}, \citenamefont {Gu{\'{e}}na}, \citenamefont {Abgrall}, \citenamefont
  {Gurov}, \citenamefont {Rosenbusch}, \citenamefont {Rovera}, \citenamefont
  {Nag{\'{o}}rny}, \citenamefont {Gartman}, \citenamefont {Westergaard},
  \citenamefont {Tobar}, \citenamefont {Lours}, \citenamefont {Santarelli},
  \citenamefont {Clairon}, \citenamefont {Bize}, \citenamefont {Laurent},
  \citenamefont {Lemonde},\ and\ \citenamefont {Lodewyck}}]{LeTargat2013}%
  \BibitemOpen
  \bibfield  {author} {\bibinfo {author} {\bibfnamefont {R.}~\bibnamefont {{Le
  Targat}}}, \bibinfo {author} {\bibfnamefont {L.}~\bibnamefont {Lorini}},
  \bibinfo {author} {\bibfnamefont {Y.}~\bibnamefont {{Le Coq}}}, \bibinfo
  {author} {\bibfnamefont {M.}~\bibnamefont {Zawada}}, \bibinfo {author}
  {\bibfnamefont {J.}~\bibnamefont {Gu{\'{e}}na}}, \bibinfo {author}
  {\bibfnamefont {M.}~\bibnamefont {Abgrall}}, \bibinfo {author} {\bibfnamefont
  {M.}~\bibnamefont {Gurov}}, \bibinfo {author} {\bibfnamefont
  {P.}~\bibnamefont {Rosenbusch}}, \bibinfo {author} {\bibfnamefont {D.~G.}\
  \bibnamefont {Rovera}}, \bibinfo {author} {\bibfnamefont {B.}~\bibnamefont
  {Nag{\'{o}}rny}}, \bibinfo {author} {\bibfnamefont {R.}~\bibnamefont
  {Gartman}}, \bibinfo {author} {\bibfnamefont {P.~G.}\ \bibnamefont
  {Westergaard}}, \bibinfo {author} {\bibfnamefont {M.~E.}\ \bibnamefont
  {Tobar}}, \bibinfo {author} {\bibfnamefont {M.}~\bibnamefont {Lours}},
  \bibinfo {author} {\bibfnamefont {G.}~\bibnamefont {Santarelli}}, \bibinfo
  {author} {\bibfnamefont {A.}~\bibnamefont {Clairon}}, \bibinfo {author}
  {\bibfnamefont {S.}~\bibnamefont {Bize}}, \bibinfo {author} {\bibfnamefont
  {P.}~\bibnamefont {Laurent}}, \bibinfo {author} {\bibfnamefont
  {P.}~\bibnamefont {Lemonde}},\ and\ \bibinfo {author} {\bibfnamefont
  {J.}~\bibnamefont {Lodewyck}},\ }\bibfield  {title} {\bibinfo {title}
  {{Experimental realization of an optical second with strontium lattice
  clocks}},\ }\href {https://doi.org/10.1038/ncomms3109} {\bibfield  {journal}
  {\bibinfo  {journal} {Nature Communications}\ }\textbf {\bibinfo {volume}
  {4}},\ \bibinfo {pages} {2109} (\bibinfo {year} {2013})}\BibitemShut
  {NoStop}%
\bibitem [{\citenamefont {Falke}\ \emph {et~al.}(2014)\citenamefont {Falke},
  \citenamefont {Lemke}, \citenamefont {Grebing}, \citenamefont {Lipphardt},
  \citenamefont {Weyers}, \citenamefont {Gerginov}, \citenamefont {Huntemann},
  \citenamefont {Hagemann}, \citenamefont {Al-Masoudi}, \citenamefont
  {Häfner}, \citenamefont {Vogt}, \citenamefont {Sterr},\ and\ \citenamefont
  {Lisdat}}]{Falke2014}%
  \BibitemOpen
  \bibfield  {author} {\bibinfo {author} {\bibfnamefont {S.}~\bibnamefont
  {Falke}}, \bibinfo {author} {\bibfnamefont {N.}~\bibnamefont {Lemke}},
  \bibinfo {author} {\bibfnamefont {C.}~\bibnamefont {Grebing}}, \bibinfo
  {author} {\bibfnamefont {B.}~\bibnamefont {Lipphardt}}, \bibinfo {author}
  {\bibfnamefont {S.}~\bibnamefont {Weyers}}, \bibinfo {author} {\bibfnamefont
  {V.}~\bibnamefont {Gerginov}}, \bibinfo {author} {\bibfnamefont
  {N.}~\bibnamefont {Huntemann}}, \bibinfo {author} {\bibfnamefont
  {C.}~\bibnamefont {Hagemann}}, \bibinfo {author} {\bibfnamefont
  {A.}~\bibnamefont {Al-Masoudi}}, \bibinfo {author} {\bibfnamefont
  {S.}~\bibnamefont {Häfner}}, \bibinfo {author} {\bibfnamefont
  {S.}~\bibnamefont {Vogt}}, \bibinfo {author} {\bibfnamefont {U.}~\bibnamefont
  {Sterr}},\ and\ \bibinfo {author} {\bibfnamefont {C.}~\bibnamefont
  {Lisdat}},\ }\bibfield  {title} {\bibinfo {title} {A strontium lattice clock
  with $3\times10^{-17}$ inaccuracy and its frequency},\ }\href
  {https://doi.org/10.1088/1367-2630/16/7/073023} {\bibfield  {journal}
  {\bibinfo  {journal} {New Journal of Physics}\ }\textbf {\bibinfo {volume}
  {16}},\ \bibinfo {pages} {073023} (\bibinfo {year} {2014})}\BibitemShut
  {NoStop}%
\bibitem [{\citenamefont {Nicholson}\ \emph {et~al.}(2015)\citenamefont
  {Nicholson}, \citenamefont {Campbell}, \citenamefont {Hutson}, \citenamefont
  {Marti}, \citenamefont {Bloom}, \citenamefont {McNally}, \citenamefont
  {Zhang}, \citenamefont {Barrett}, \citenamefont {Safronova}, \citenamefont
  {Strouse}, \citenamefont {Tew},\ and\ \citenamefont {Ye}}]{Nicholson2015a}%
  \BibitemOpen
  \bibfield  {author} {\bibinfo {author} {\bibfnamefont {T.}~\bibnamefont
  {Nicholson}}, \bibinfo {author} {\bibfnamefont {S.}~\bibnamefont {Campbell}},
  \bibinfo {author} {\bibfnamefont {R.}~\bibnamefont {Hutson}}, \bibinfo
  {author} {\bibfnamefont {G.}~\bibnamefont {Marti}}, \bibinfo {author}
  {\bibfnamefont {B.}~\bibnamefont {Bloom}}, \bibinfo {author} {\bibfnamefont
  {R.}~\bibnamefont {McNally}}, \bibinfo {author} {\bibfnamefont
  {W.}~\bibnamefont {Zhang}}, \bibinfo {author} {\bibfnamefont
  {M.}~\bibnamefont {Barrett}}, \bibinfo {author} {\bibfnamefont
  {M.}~\bibnamefont {Safronova}}, \bibinfo {author} {\bibfnamefont
  {G.}~\bibnamefont {Strouse}}, \bibinfo {author} {\bibfnamefont
  {W.}~\bibnamefont {Tew}},\ and\ \bibinfo {author} {\bibfnamefont
  {J.}~\bibnamefont {Ye}},\ }\bibfield  {title} {\bibinfo {title} {{Systematic
  evaluation of an atomic clock at $2\times10^{-18}$ total uncertainty}},\
  }\href {https://doi.org/10.1038/ncomms7896} {\bibfield  {journal} {\bibinfo
  {journal} {Nature Communications}\ }\textbf {\bibinfo {volume} {6}},\
  \bibinfo {pages} {6896} (\bibinfo {year} {2015})}\BibitemShut {NoStop}%
\bibitem [{\citenamefont {Koller}\ \emph {et~al.}(2017)\citenamefont {Koller},
  \citenamefont {Grotti}, \citenamefont {Vogt}, \citenamefont {Al-Masoudi},
  \citenamefont {D\"orscher}, \citenamefont {H\"afner}, \citenamefont {Sterr},\
  and\ \citenamefont {Lisdat}}]{Koller2017}%
  \BibitemOpen
  \bibfield  {author} {\bibinfo {author} {\bibfnamefont {S.~B.}\ \bibnamefont
  {Koller}}, \bibinfo {author} {\bibfnamefont {J.}~\bibnamefont {Grotti}},
  \bibinfo {author} {\bibfnamefont {S.}~\bibnamefont {Vogt}}, \bibinfo {author}
  {\bibfnamefont {A.}~\bibnamefont {Al-Masoudi}}, \bibinfo {author}
  {\bibfnamefont {S.}~\bibnamefont {D\"orscher}}, \bibinfo {author}
  {\bibfnamefont {S.}~\bibnamefont {H\"afner}}, \bibinfo {author}
  {\bibfnamefont {U.}~\bibnamefont {Sterr}},\ and\ \bibinfo {author}
  {\bibfnamefont {C.}~\bibnamefont {Lisdat}},\ }\bibfield  {title} {\bibinfo
  {title} {Transportable optical lattice clock with
  $7\ifmmode\times\else\texttimes\fi{}{10}^{\ensuremath{-}17}$ uncertainty},\
  }\href {https://doi.org/10.1103/PhysRevLett.118.073601} {\bibfield  {journal}
  {\bibinfo  {journal} {Phys. Rev. Lett.}\ }\textbf {\bibinfo {volume} {118}},\
  \bibinfo {pages} {073601} (\bibinfo {year} {2017})}\BibitemShut {NoStop}%
\bibitem [{\citenamefont {Hisai}\ \emph {et~al.}(2021)\citenamefont {Hisai},
  \citenamefont {Akamatsu}, \citenamefont {Kobayashi}, \citenamefont {Hosaka},
  \citenamefont {Inaba}, \citenamefont {Hong},\ and\ \citenamefont
  {Yasuda}}]{Hisai2021}%
  \BibitemOpen
  \bibfield  {author} {\bibinfo {author} {\bibfnamefont {Y.}~\bibnamefont
  {Hisai}}, \bibinfo {author} {\bibfnamefont {D.}~\bibnamefont {Akamatsu}},
  \bibinfo {author} {\bibfnamefont {T.}~\bibnamefont {Kobayashi}}, \bibinfo
  {author} {\bibfnamefont {K.}~\bibnamefont {Hosaka}}, \bibinfo {author}
  {\bibfnamefont {H.}~\bibnamefont {Inaba}}, \bibinfo {author} {\bibfnamefont
  {F.-L.}\ \bibnamefont {Hong}},\ and\ \bibinfo {author} {\bibfnamefont
  {M.}~\bibnamefont {Yasuda}},\ }\bibfield  {title} {\bibinfo {title}
  {{{Improved frequency ratio measurement with $^{87}$Sr and $^{171}$Yb optical
  lattice clocks at NMIJ}}},\ }\href {https://doi.org/10.1088/1681-7575/abc104}
  {\bibfield  {journal} {\bibinfo  {journal} {Metrologia}\ }\textbf {\bibinfo
  {volume} {58}},\ \bibinfo {pages} {015008} (\bibinfo {year}
  {2021})}\BibitemShut {NoStop}%
\bibitem [{\citenamefont {Ohmae}\ \emph {et~al.}(2021)\citenamefont {Ohmae},
  \citenamefont {Takamoto}, \citenamefont {Takahashi}, \citenamefont {Kokubun},
  \citenamefont {Araki}, \citenamefont {Hinton}, \citenamefont {Ushijima},
  \citenamefont {Muramatsu}, \citenamefont {Furumiya}, \citenamefont {Sakai},
  \citenamefont {Moriya}, \citenamefont {Kamiya}, \citenamefont {Fujii},
  \citenamefont {Muramatsu}, \citenamefont {Shiimado},\ and\ \citenamefont
  {Katori}}]{Ohmae2021}%
  \BibitemOpen
  \bibfield  {author} {\bibinfo {author} {\bibfnamefont {N.}~\bibnamefont
  {Ohmae}}, \bibinfo {author} {\bibfnamefont {M.}~\bibnamefont {Takamoto}},
  \bibinfo {author} {\bibfnamefont {Y.}~\bibnamefont {Takahashi}}, \bibinfo
  {author} {\bibfnamefont {M.}~\bibnamefont {Kokubun}}, \bibinfo {author}
  {\bibfnamefont {K.}~\bibnamefont {Araki}}, \bibinfo {author} {\bibfnamefont
  {A.}~\bibnamefont {Hinton}}, \bibinfo {author} {\bibfnamefont
  {I.}~\bibnamefont {Ushijima}}, \bibinfo {author} {\bibfnamefont
  {T.}~\bibnamefont {Muramatsu}}, \bibinfo {author} {\bibfnamefont
  {T.}~\bibnamefont {Furumiya}}, \bibinfo {author} {\bibfnamefont
  {Y.}~\bibnamefont {Sakai}}, \bibinfo {author} {\bibfnamefont
  {N.}~\bibnamefont {Moriya}}, \bibinfo {author} {\bibfnamefont
  {N.}~\bibnamefont {Kamiya}}, \bibinfo {author} {\bibfnamefont
  {K.}~\bibnamefont {Fujii}}, \bibinfo {author} {\bibfnamefont
  {R.}~\bibnamefont {Muramatsu}}, \bibinfo {author} {\bibfnamefont
  {T.}~\bibnamefont {Shiimado}},\ and\ \bibinfo {author} {\bibfnamefont
  {H.}~\bibnamefont {Katori}},\ }\bibfield  {title} {\bibinfo {title}
  {{Transportable Strontium Optical Lattice Clocks Operated Outside Laboratory
  at the Level of $10^{-18}$ Uncertainty}},\ }\href
  {https://doi.org/10.1002/qute.202100015} {\bibfield  {journal} {\bibinfo
  {journal} {Advanced Quantum Technologies}\ }\textbf {\bibinfo {volume} {4}},\
  \bibinfo {pages} {2100015} (\bibinfo {year} {2021})}\BibitemShut {NoStop}%
\bibitem [{\citenamefont {Ushijima}\ \emph {et~al.}(2015)\citenamefont
  {Ushijima}, \citenamefont {Takamoto}, \citenamefont {Das}, \citenamefont
  {Ohkubo},\ and\ \citenamefont {Katori}}]{Ushijima2015}%
  \BibitemOpen
  \bibfield  {author} {\bibinfo {author} {\bibfnamefont {I.}~\bibnamefont
  {Ushijima}}, \bibinfo {author} {\bibfnamefont {M.}~\bibnamefont {Takamoto}},
  \bibinfo {author} {\bibfnamefont {M.}~\bibnamefont {Das}}, \bibinfo {author}
  {\bibfnamefont {T.}~\bibnamefont {Ohkubo}},\ and\ \bibinfo {author}
  {\bibfnamefont {H.}~\bibnamefont {Katori}},\ }\bibfield  {title} {\bibinfo
  {title} {Cryogenic optical lattice clocks},\ }\href
  {https://doi.org/10.1038/nphoton.2015.5} {\bibfield  {journal} {\bibinfo
  {journal} {Nature Photonics}\ }\textbf {\bibinfo {volume} {9}},\ \bibinfo
  {pages} {185} (\bibinfo {year} {2015})}\BibitemShut {NoStop}%
\bibitem [{\citenamefont {Ablewski}\ \emph {et~al.}(2020)\citenamefont
  {Ablewski}, \citenamefont {Bober},\ and\ \citenamefont
  {Zawada}}]{Ablewski2020}%
  \BibitemOpen
  \bibfield  {author} {\bibinfo {author} {\bibfnamefont {P.}~\bibnamefont
  {Ablewski}}, \bibinfo {author} {\bibfnamefont {M.}~\bibnamefont {Bober}},\
  and\ \bibinfo {author} {\bibfnamefont {M.}~\bibnamefont {Zawada}},\
  }\bibfield  {title} {\bibinfo {title} {Emissivities of vacuum compatible
  materials: towards minimising blackbody radiation shift uncertainty in
  optical atomic clocks at room temperatures},\ }\href
  {https://doi.org/10.1088/1681-7575/ab63ae} {\bibfield  {journal} {\bibinfo
  {journal} {Metrologia}\ }\textbf {\bibinfo {volume} {57}},\ \bibinfo {pages}
  {035004} (\bibinfo {year} {2020})}\BibitemShut {NoStop}%
\bibitem [{\citenamefont {Yudin}\ \emph {et~al.}(2021)\citenamefont {Yudin},
  \citenamefont {Taichenachev}, \citenamefont {Basalaev}, \citenamefont
  {Prudnikov}, \citenamefont {Fürst}, \citenamefont {Mehlstäubler},\ and\
  \citenamefont {Bagayev}}]{Yudin2021}%
  \BibitemOpen
  \bibfield  {author} {\bibinfo {author} {\bibfnamefont {V.~I.}\ \bibnamefont
  {Yudin}}, \bibinfo {author} {\bibfnamefont {A.~V.}\ \bibnamefont
  {Taichenachev}}, \bibinfo {author} {\bibfnamefont {M.~Y.}\ \bibnamefont
  {Basalaev}}, \bibinfo {author} {\bibfnamefont {O.~N.}\ \bibnamefont
  {Prudnikov}}, \bibinfo {author} {\bibfnamefont {H.~A.}\ \bibnamefont
  {Fürst}}, \bibinfo {author} {\bibfnamefont {T.~E.}\ \bibnamefont
  {Mehlstäubler}},\ and\ \bibinfo {author} {\bibfnamefont {S.~N.}\
  \bibnamefont {Bagayev}},\ }\bibfield  {title} {\bibinfo {title} {Combined
  atomic clock with blackbody-radiation-shift-induced instability below
  10$^{-19}$ under natural environment conditions},\ }\href
  {https://doi.org/10.1088/1367-2630/abe160} {\bibfield  {journal} {\bibinfo
  {journal} {New Journal of Physics}\ }\textbf {\bibinfo {volume} {23}},\
  \bibinfo {pages} {023032} (\bibinfo {year} {2021})}\BibitemShut {NoStop}%
\bibitem [{\citenamefont {Middelmann}\ \emph {et~al.}(2011)\citenamefont
  {Middelmann}, \citenamefont {Lisdat}, \citenamefont {Falke}, \citenamefont
  {Vellore~Winfred}, \citenamefont {Riehle},\ and\ \citenamefont
  {Sterr}}]{Middelmann2011}%
  \BibitemOpen
  \bibfield  {author} {\bibinfo {author} {\bibfnamefont {T.}~\bibnamefont
  {Middelmann}}, \bibinfo {author} {\bibfnamefont {C.}~\bibnamefont {Lisdat}},
  \bibinfo {author} {\bibfnamefont {S.}~\bibnamefont {Falke}}, \bibinfo
  {author} {\bibfnamefont {J.~S.~R.}\ \bibnamefont {Vellore~Winfred}}, \bibinfo
  {author} {\bibfnamefont {F.}~\bibnamefont {Riehle}},\ and\ \bibinfo {author}
  {\bibfnamefont {U.}~\bibnamefont {Sterr}},\ }\bibfield  {title} {\bibinfo
  {title} {Tackling the blackbody shift in a strontium optical lattice clock},\
  }\href {https://doi.org/10.1109/TIM.2010.2088470} {\bibfield  {journal}
  {\bibinfo  {journal} {IEEE Transactions on Instrumentation and Measurement}\
  }\textbf {\bibinfo {volume} {60}},\ \bibinfo {pages} {2550} (\bibinfo {year}
  {2011})}\BibitemShut {NoStop}%
\bibitem [{\citenamefont {Middelmann}\ \emph {et~al.}(2012)\citenamefont
  {Middelmann}, \citenamefont {Falke}, \citenamefont {Lisdat},\ and\
  \citenamefont {Sterr}}]{Middelmann2012}%
  \BibitemOpen
  \bibfield  {author} {\bibinfo {author} {\bibfnamefont {T.}~\bibnamefont
  {Middelmann}}, \bibinfo {author} {\bibfnamefont {S.}~\bibnamefont {Falke}},
  \bibinfo {author} {\bibfnamefont {C.}~\bibnamefont {Lisdat}},\ and\ \bibinfo
  {author} {\bibfnamefont {U.}~\bibnamefont {Sterr}},\ }\bibfield  {title}
  {\bibinfo {title} {High accuracy correction of blackbody radiation shift in
  an optical lattice clock},\ }\href
  {https://doi.org/10.1103/PhysRevLett.109.263004} {\bibfield  {journal}
  {\bibinfo  {journal} {Phys. Rev. Lett.}\ }\textbf {\bibinfo {volume} {109}},\
  \bibinfo {pages} {263004} (\bibinfo {year} {2012})}\BibitemShut {NoStop}%
\bibitem [{\citenamefont {Lisdat}\ \emph {et~al.}(2021)\citenamefont {Lisdat},
  \citenamefont {D\"orscher}, \citenamefont {Nosske},\ and\ \citenamefont
  {Sterr}}]{Lisdat2021}%
  \BibitemOpen
  \bibfield  {author} {\bibinfo {author} {\bibfnamefont {C.}~\bibnamefont
  {Lisdat}}, \bibinfo {author} {\bibfnamefont {S.}~\bibnamefont {D\"orscher}},
  \bibinfo {author} {\bibfnamefont {I.}~\bibnamefont {Nosske}},\ and\ \bibinfo
  {author} {\bibfnamefont {U.}~\bibnamefont {Sterr}},\ }\bibfield  {title}
  {\bibinfo {title} {Blackbody radiation shift in strontium lattice clocks
  revisited},\ }\href {https://doi.org/10.1103/PhysRevResearch.3.L042036}
  {\bibfield  {journal} {\bibinfo  {journal} {Phys. Rev. Res.}\ }\textbf
  {\bibinfo {volume} {3}},\ \bibinfo {pages} {L042036} (\bibinfo {year}
  {2021})}\BibitemShut {NoStop}%
\bibitem [{\citenamefont {Porsev}\ and\ \citenamefont
  {Derevianko}(2006)}]{Porsev2006}%
  \BibitemOpen
  \bibfield  {author} {\bibinfo {author} {\bibfnamefont {S.~G.}\ \bibnamefont
  {Porsev}}\ and\ \bibinfo {author} {\bibfnamefont {A.}~\bibnamefont
  {Derevianko}},\ }\bibfield  {title} {\bibinfo {title} {Multipolar theory of
  blackbody radiation shift of atomic energy levels and its implications for
  optical lattice clocks},\ }\href {https://doi.org/10.1103/PhysRevA.74.020502}
  {\bibfield  {journal} {\bibinfo  {journal} {Phys. Rev. A}\ }\textbf {\bibinfo
  {volume} {74}},\ \bibinfo {pages} {020502} (\bibinfo {year}
  {2006})}\BibitemShut {NoStop}%
\bibitem [{\citenamefont {Safronova}\ \emph {et~al.}(2013)\citenamefont
  {Safronova}, \citenamefont {Porsev}, \citenamefont {Safronova}, \citenamefont
  {Kozlov},\ and\ \citenamefont {Clark}}]{Safronova2013}%
  \BibitemOpen
  \bibfield  {author} {\bibinfo {author} {\bibfnamefont {M.~S.}\ \bibnamefont
  {Safronova}}, \bibinfo {author} {\bibfnamefont {S.~G.}\ \bibnamefont
  {Porsev}}, \bibinfo {author} {\bibfnamefont {U.~I.}\ \bibnamefont
  {Safronova}}, \bibinfo {author} {\bibfnamefont {M.~G.}\ \bibnamefont
  {Kozlov}},\ and\ \bibinfo {author} {\bibfnamefont {C.~W.}\ \bibnamefont
  {Clark}},\ }\bibfield  {title} {\bibinfo {title} {Blackbody-radiation shift
  in the {S}r optical atomic clock},\ }\href
  {https://doi.org/10.1103/PHYSREVA.87.012509/FIGURES/1/MEDIUM} {\bibfield
  {journal} {\bibinfo  {journal} {Physical Review A}\ }\textbf {\bibinfo
  {volume} {87}},\ \bibinfo {pages} {012509} (\bibinfo {year}
  {2013})}\BibitemShut {NoStop}%
\bibitem [{\citenamefont {Leung}\ \emph {et~al.}(2023)\citenamefont {Leung},
  \citenamefont {Iritani}, \citenamefont {Tiberi}, \citenamefont {Majewska},
  \citenamefont {Borkowski}, \citenamefont {Moszynski},\ and\ \citenamefont
  {Zelevinsky}}]{Leung2023}%
  \BibitemOpen
  \bibfield  {author} {\bibinfo {author} {\bibfnamefont {K.~H.}\ \bibnamefont
  {Leung}}, \bibinfo {author} {\bibfnamefont {B.}~\bibnamefont {Iritani}},
  \bibinfo {author} {\bibfnamefont {E.}~\bibnamefont {Tiberi}}, \bibinfo
  {author} {\bibfnamefont {I.}~\bibnamefont {Majewska}}, \bibinfo {author}
  {\bibfnamefont {M.}~\bibnamefont {Borkowski}}, \bibinfo {author}
  {\bibfnamefont {R.}~\bibnamefont {Moszynski}},\ and\ \bibinfo {author}
  {\bibfnamefont {T.}~\bibnamefont {Zelevinsky}},\ }\bibfield  {title}
  {\bibinfo {title} {Terahertz vibrational molecular clock with systematic
  uncertainty at the ${10}^{\ensuremath{-}14}$ level},\ }\href
  {https://doi.org/10.1103/PhysRevX.13.011047} {\bibfield  {journal} {\bibinfo
  {journal} {Phys. Rev. X}\ }\textbf {\bibinfo {volume} {13}},\ \bibinfo
  {pages} {011047} (\bibinfo {year} {2023})}\BibitemShut {NoStop}%
\bibitem [{\citenamefont {Leung}\ \emph {et~al.}(2020)\citenamefont {Leung},
  \citenamefont {Majewska}, \citenamefont {Bekker}, \citenamefont {Lee},
  \citenamefont {Tiberi}, \citenamefont {Kondov}, \citenamefont {Moszynski},\
  and\ \citenamefont {Zelevinsky}}]{Leung2020}%
  \BibitemOpen
  \bibfield  {author} {\bibinfo {author} {\bibfnamefont {K.~H.}\ \bibnamefont
  {Leung}}, \bibinfo {author} {\bibfnamefont {I.}~\bibnamefont {Majewska}},
  \bibinfo {author} {\bibfnamefont {H.}~\bibnamefont {Bekker}}, \bibinfo
  {author} {\bibfnamefont {C.-H.}\ \bibnamefont {Lee}}, \bibinfo {author}
  {\bibfnamefont {E.}~\bibnamefont {Tiberi}}, \bibinfo {author} {\bibfnamefont
  {S.~S.}\ \bibnamefont {Kondov}}, \bibinfo {author} {\bibfnamefont
  {R.}~\bibnamefont {Moszynski}},\ and\ \bibinfo {author} {\bibfnamefont
  {T.}~\bibnamefont {Zelevinsky}},\ }\bibfield  {title} {\bibinfo {title}
  {Transition strength measurements to guide magic wavelength selection in
  optically trapped molecules},\ }\href
  {https://doi.org/10.1103/PhysRevLett.125.153001} {\bibfield  {journal}
  {\bibinfo  {journal} {Phys. Rev. Lett.}\ }\textbf {\bibinfo {volume} {125}},\
  \bibinfo {pages} {153001} (\bibinfo {year} {2020})}\BibitemShut {NoStop}%
\bibitem [{\citenamefont {Leung}\ \emph {et~al.}(2021)\citenamefont {Leung},
  \citenamefont {Tiberi}, \citenamefont {Iritani}, \citenamefont {Majewska},
  \citenamefont {Moszynski},\ and\ \citenamefont {Zelevinsky}}]{Leung2021}%
  \BibitemOpen
  \bibfield  {author} {\bibinfo {author} {\bibfnamefont {K.~H.}\ \bibnamefont
  {Leung}}, \bibinfo {author} {\bibfnamefont {E.}~\bibnamefont {Tiberi}},
  \bibinfo {author} {\bibfnamefont {B.}~\bibnamefont {Iritani}}, \bibinfo
  {author} {\bibfnamefont {I.}~\bibnamefont {Majewska}}, \bibinfo {author}
  {\bibfnamefont {R.}~\bibnamefont {Moszynski}},\ and\ \bibinfo {author}
  {\bibfnamefont {T.}~\bibnamefont {Zelevinsky}},\ }\bibfield  {title}
  {\bibinfo {title} {{Ultracold $^{88}$Sr$_2$ molecules in the absolute ground
  state}},\ }\href {https://doi.org/10.1088/1367-2630/ac2dac} {\bibfield
  {journal} {\bibinfo  {journal} {New Journal of Physics}\ }\textbf {\bibinfo
  {volume} {23}},\ \bibinfo {pages} {115002} (\bibinfo {year}
  {2021})}\BibitemShut {NoStop}%
\bibitem [{\citenamefont {Zelevinsky}\ \emph {et~al.}(2006)\citenamefont
  {Zelevinsky}, \citenamefont {Boyd}, \citenamefont {Ludlow}, \citenamefont
  {Ido}, \citenamefont {Ye}, \citenamefont {Ciury{\l}o}, \citenamefont
  {Naidon},\ and\ \citenamefont {Julienne}}]{Zelevinsky2006}%
  \BibitemOpen
  \bibfield  {author} {\bibinfo {author} {\bibfnamefont {T.}~\bibnamefont
  {Zelevinsky}}, \bibinfo {author} {\bibfnamefont {M.~M.}\ \bibnamefont
  {Boyd}}, \bibinfo {author} {\bibfnamefont {A.~D.}\ \bibnamefont {Ludlow}},
  \bibinfo {author} {\bibfnamefont {T.}~\bibnamefont {Ido}}, \bibinfo {author}
  {\bibfnamefont {J.}~\bibnamefont {Ye}}, \bibinfo {author} {\bibfnamefont
  {R.}~\bibnamefont {Ciury{\l}o}}, \bibinfo {author} {\bibfnamefont
  {P.}~\bibnamefont {Naidon}},\ and\ \bibinfo {author} {\bibfnamefont {P.~S.}\
  \bibnamefont {Julienne}},\ }\bibfield  {title} {\bibinfo {title} {{Narrow
  Line Photoassociation in an Optical Lattice}},\ }\href
  {https://doi.org/10.1103/PhysRevLett.96.203201} {\bibfield  {journal}
  {\bibinfo  {journal} {Physical Review Letters}\ }\textbf {\bibinfo {volume}
  {96}},\ \bibinfo {pages} {203201} (\bibinfo {year} {2006})}\BibitemShut
  {NoStop}%
\bibitem [{\citenamefont {McGuyer}\ \emph {et~al.}(2015)\citenamefont
  {McGuyer}, \citenamefont {McDonald}, \citenamefont {Iwata}, \citenamefont
  {Tarallo}, \citenamefont {Grier}, \citenamefont {Apfelbeck},\ and\
  \citenamefont {Zelevinsky}}]{McGuyer2015a}%
  \BibitemOpen
  \bibfield  {author} {\bibinfo {author} {\bibfnamefont {B.~H.}\ \bibnamefont
  {McGuyer}}, \bibinfo {author} {\bibfnamefont {M.}~\bibnamefont {McDonald}},
  \bibinfo {author} {\bibfnamefont {G.~Z.}\ \bibnamefont {Iwata}}, \bibinfo
  {author} {\bibfnamefont {M.~G.}\ \bibnamefont {Tarallo}}, \bibinfo {author}
  {\bibfnamefont {A.~T.}\ \bibnamefont {Grier}}, \bibinfo {author}
  {\bibfnamefont {F.}~\bibnamefont {Apfelbeck}},\ and\ \bibinfo {author}
  {\bibfnamefont {T.}~\bibnamefont {Zelevinsky}},\ }\bibfield  {title}
  {\bibinfo {title} {{High-precision spectroscopy of ultracold molecules in an
  optical lattice}},\ }\href {https://doi.org/10.1088/1367-2630/17/5/055004}
  {\bibfield  {journal} {\bibinfo  {journal} {New J. Phys}\ }\textbf {\bibinfo
  {volume} {17}},\ \bibinfo {pages} {055004} (\bibinfo {year}
  {2015})}\BibitemShut {NoStop}%
\bibitem [{\citenamefont {Autler}\ and\ \citenamefont
  {Townes}(1955)}]{autler1955}%
  \BibitemOpen
  \bibfield  {author} {\bibinfo {author} {\bibfnamefont {S.~H.}\ \bibnamefont
  {Autler}}\ and\ \bibinfo {author} {\bibfnamefont {C.~H.}\ \bibnamefont
  {Townes}},\ }\bibfield  {title} {\bibinfo {title} {Stark effect in rapidly
  varying fields},\ }\href {https://doi.org/10.1103/PhysRev.100.703} {\bibfield
   {journal} {\bibinfo  {journal} {Phys. Rev.}\ }\textbf {\bibinfo {volume}
  {100}},\ \bibinfo {pages} {703} (\bibinfo {year} {1955})}\BibitemShut
  {NoStop}%
\bibitem [{\citenamefont {Townes}\ and\ \citenamefont
  {Schawlow}(1975)}]{townes2013microwave}%
  \BibitemOpen
  \bibfield  {author} {\bibinfo {author} {\bibfnamefont {C.~H.}\ \bibnamefont
  {Townes}}\ and\ \bibinfo {author} {\bibfnamefont {A.~L.}\ \bibnamefont
  {Schawlow}},\ }\href@noop {} {\emph {\bibinfo {title} {Microwave
  spectroscopy}}}\ (\bibinfo  {publisher} {Dover Publications},\ \bibinfo
  {year} {1975})\BibitemShut {NoStop}%
\bibitem [{\citenamefont {Jones}\ \emph {et~al.}(2006)\citenamefont {Jones},
  \citenamefont {Tiesinga}, \citenamefont {Lett},\ and\ \citenamefont
  {Julienne}}]{Jones2006}%
  \BibitemOpen
  \bibfield  {author} {\bibinfo {author} {\bibfnamefont {K.~M.}\ \bibnamefont
  {Jones}}, \bibinfo {author} {\bibfnamefont {E.}~\bibnamefont {Tiesinga}},
  \bibinfo {author} {\bibfnamefont {P.~D.}\ \bibnamefont {Lett}},\ and\
  \bibinfo {author} {\bibfnamefont {P.~S.}\ \bibnamefont {Julienne}},\
  }\bibfield  {title} {\bibinfo {title} {{Ultracold photoassociation
  spectroscopy: Long-range molecules and atomic scattering}},\ }\href
  {https://doi.org/10.1103/RevModPhys.78.483} {\bibfield  {journal} {\bibinfo
  {journal} {Reviews of Modern Physics}\ }\textbf {\bibinfo {volume} {78}},\
  \bibinfo {pages} {483} (\bibinfo {year} {2006})}\BibitemShut {NoStop}%
\bibitem [{\citenamefont {Martinez~de Escobar}\ \emph
  {et~al.}(2008)\citenamefont {Martinez~de Escobar}, \citenamefont {Mickelson},
  \citenamefont {Pellegrini}, \citenamefont {Nagel}, \citenamefont {Traverso},
  \citenamefont {Yan}, \citenamefont {C\^ot\'e},\ and\ \citenamefont
  {Killian}}]{martinez2008}%
  \BibitemOpen
  \bibfield  {author} {\bibinfo {author} {\bibfnamefont {Y.~N.}\ \bibnamefont
  {Martinez~de Escobar}}, \bibinfo {author} {\bibfnamefont {P.~G.}\
  \bibnamefont {Mickelson}}, \bibinfo {author} {\bibfnamefont {P.}~\bibnamefont
  {Pellegrini}}, \bibinfo {author} {\bibfnamefont {S.~B.}\ \bibnamefont
  {Nagel}}, \bibinfo {author} {\bibfnamefont {A.}~\bibnamefont {Traverso}},
  \bibinfo {author} {\bibfnamefont {M.}~\bibnamefont {Yan}}, \bibinfo {author}
  {\bibfnamefont {R.}~\bibnamefont {C\^ot\'e}},\ and\ \bibinfo {author}
  {\bibfnamefont {T.~C.}\ \bibnamefont {Killian}},\ }\bibfield  {title}
  {\bibinfo {title} {Two-photon photoassociative spectroscopy of ultracold
  $^{88}\mathrm{Sr}$},\ }\href {https://doi.org/10.1103/PhysRevA.78.062708}
  {\bibfield  {journal} {\bibinfo  {journal} {Phys. Rev. A}\ }\textbf {\bibinfo
  {volume} {78}},\ \bibinfo {pages} {062708} (\bibinfo {year}
  {2008})}\BibitemShut {NoStop}%
\bibitem [{\citenamefont {Kitagawa}\ \emph {et~al.}(2008)\citenamefont
  {Kitagawa}, \citenamefont {Enomoto}, \citenamefont {Kasa}, \citenamefont
  {Takahashi}, \citenamefont {Ciury{\l}o}, \citenamefont {Naidon},\ and\
  \citenamefont {Julienne}}]{Kitagawa2008}%
  \BibitemOpen
  \bibfield  {author} {\bibinfo {author} {\bibfnamefont {M.}~\bibnamefont
  {Kitagawa}}, \bibinfo {author} {\bibfnamefont {K.}~\bibnamefont {Enomoto}},
  \bibinfo {author} {\bibfnamefont {K.}~\bibnamefont {Kasa}}, \bibinfo {author}
  {\bibfnamefont {Y.}~\bibnamefont {Takahashi}}, \bibinfo {author}
  {\bibfnamefont {R.}~\bibnamefont {Ciury{\l}o}}, \bibinfo {author}
  {\bibfnamefont {P.}~\bibnamefont {Naidon}},\ and\ \bibinfo {author}
  {\bibfnamefont {P.~S.}\ \bibnamefont {Julienne}},\ }\bibfield  {title}
  {\bibinfo {title} {{Two-color photoassociation spectroscopy of ytterbium
  atoms and the precise determinations of s-wave scattering lengths}},\ }\href
  {https://doi.org/10.1103/PhysRevA.77.012719} {\bibfield  {journal} {\bibinfo
  {journal} {Physical Review A}\ }\textbf {\bibinfo {volume} {77}},\ \bibinfo
  {pages} {012719} (\bibinfo {year} {2008})}\BibitemShut {NoStop}%
\bibitem [{\citenamefont {Tiesinga}\ \emph {et~al.}(2021)\citenamefont
  {Tiesinga}, \citenamefont {Mohr}, \citenamefont {Newell},\ and\ \citenamefont
  {Taylor}}]{Tiesinga2021}%
  \BibitemOpen
  \bibfield  {author} {\bibinfo {author} {\bibfnamefont {E.}~\bibnamefont
  {Tiesinga}}, \bibinfo {author} {\bibfnamefont {P.~J.}\ \bibnamefont {Mohr}},
  \bibinfo {author} {\bibfnamefont {D.~B.}\ \bibnamefont {Newell}},\ and\
  \bibinfo {author} {\bibfnamefont {B.~N.}\ \bibnamefont {Taylor}},\ }\bibfield
   {title} {\bibinfo {title} {Codata recommended values of the fundamental
  physical constants: 2018},\ }\href
  {https://doi.org/10.1103/RevModPhys.93.025010} {\bibfield  {journal}
  {\bibinfo  {journal} {Rev. Mod. Phys.}\ }\textbf {\bibinfo {volume} {93}},\
  \bibinfo {pages} {025010} (\bibinfo {year} {2021})}\BibitemShut {NoStop}%
\bibitem [{\citenamefont {Pound}(1946)}]{Pound1946}%
  \BibitemOpen
  \bibfield  {author} {\bibinfo {author} {\bibfnamefont {R.~V.}\ \bibnamefont
  {Pound}},\ }\bibfield  {title} {\bibinfo {title} {Electronic frequency
  stabilization of microwave oscillators},\ }\href
  {https://doi.org/10.1063/1.1770414} {\bibfield  {journal} {\bibinfo
  {journal} {Review of Scientific Instruments}\ }\textbf {\bibinfo {volume}
  {17}},\ \bibinfo {pages} {490} (\bibinfo {year} {1946})}\BibitemShut
  {NoStop}%
\bibitem [{\citenamefont {Drever}\ \emph {et~al.}(1983)\citenamefont {Drever},
  \citenamefont {Hall}, \citenamefont {Kowalski}, \citenamefont {Hough},
  \citenamefont {Ford}, \citenamefont {Munley},\ and\ \citenamefont
  {Ward}}]{Drever1983}%
  \BibitemOpen
  \bibfield  {author} {\bibinfo {author} {\bibfnamefont {R.~W.}\ \bibnamefont
  {Drever}}, \bibinfo {author} {\bibfnamefont {J.~L.}\ \bibnamefont {Hall}},
  \bibinfo {author} {\bibfnamefont {F.~V.}\ \bibnamefont {Kowalski}}, \bibinfo
  {author} {\bibfnamefont {J.}~\bibnamefont {Hough}}, \bibinfo {author}
  {\bibfnamefont {G.}~\bibnamefont {Ford}}, \bibinfo {author} {\bibfnamefont
  {A.}~\bibnamefont {Munley}},\ and\ \bibinfo {author} {\bibfnamefont
  {H.}~\bibnamefont {Ward}},\ }\bibfield  {title} {\bibinfo {title} {Laser
  phase and frequency stabilization using an optical resonator},\ }\href
  {https://doi.org/10.1007/BF00702605} {\bibfield  {journal} {\bibinfo
  {journal} {Applied Physics B}\ }\textbf {\bibinfo {volume} {31}},\ \bibinfo
  {pages} {97} (\bibinfo {year} {1983})}\BibitemShut {NoStop}%
\bibitem [{\citenamefont {Safronova}(2023)}]{SafronovaPrivateCommunication}%
  \BibitemOpen
  \bibfield  {author} {\bibinfo {author} {\bibfnamefont {M.~S.}\ \bibnamefont
  {Safronova}},\ }\href@noop {} {}\bibinfo {howpublished} {{private
  communication}} (\bibinfo {year} {2023})\BibitemShut {NoStop}%
\bibitem [{\citenamefont {McDonald}\ \emph {et~al.}(2015)\citenamefont
  {McDonald}, \citenamefont {McGuyer}, \citenamefont {Iwata},\ and\
  \citenamefont {Zelevinsky}}]{McDonald2015}%
  \BibitemOpen
  \bibfield  {author} {\bibinfo {author} {\bibfnamefont {M.}~\bibnamefont
  {McDonald}}, \bibinfo {author} {\bibfnamefont {B.~H.}\ \bibnamefont
  {McGuyer}}, \bibinfo {author} {\bibfnamefont {G.~Z.}\ \bibnamefont {Iwata}},\
  and\ \bibinfo {author} {\bibfnamefont {T.}~\bibnamefont {Zelevinsky}},\
  }\bibfield  {title} {\bibinfo {title} {Thermometry via light shifts in
  optical lattices},\ }\href {https://doi.org/10.1103/PhysRevLett.114.023001}
  {\bibfield  {journal} {\bibinfo  {journal} {Phys. Rev. Lett.}\ }\textbf
  {\bibinfo {volume} {114}},\ \bibinfo {pages} {023001} (\bibinfo {year}
  {2015})}\BibitemShut {NoStop}%
\bibitem [{\citenamefont {{Le Roy}}\ and\ \citenamefont
  {Bernstein}(1970)}]{Leroy1970}%
  \BibitemOpen
  \bibfield  {author} {\bibinfo {author} {\bibfnamefont {R.~J.}\ \bibnamefont
  {{Le Roy}}}\ and\ \bibinfo {author} {\bibfnamefont {R.~B.}\ \bibnamefont
  {Bernstein}},\ }\bibfield  {title} {\bibinfo {title} {{Dissociation Energy
  and Long-Range Potential of Diatomic Molecules from Vibrational Spacings of
  Higher Levels}},\ }\href {https://doi.org/10.1063/1.1697142} {\bibfield
  {journal} {\bibinfo  {journal} {The Journal of Chemical Physics}\ }\textbf
  {\bibinfo {volume} {52}},\ \bibinfo {pages} {3869} (\bibinfo {year}
  {1970})}\BibitemShut {NoStop}%
\bibitem [{\citenamefont {{Le Roy}}(1973)}]{LeRoy1973}%
  \BibitemOpen
  \bibfield  {author} {\bibinfo {author} {\bibfnamefont {R.~J.}\ \bibnamefont
  {{Le Roy}}},\ }\href {https://doi.org/10.1039/9781847556684-00113} {\emph
  {\bibinfo {title} {Molecular Spectroscopy - Volume I, A Specialist Periodical
  Report of the Chemical Society}}}\ (\bibinfo  {publisher} {The Chemical
  Society, London},\ \bibinfo {year} {1973})\ pp.\ \bibinfo {pages}
  {113--171}\BibitemShut {NoStop}%
\bibitem [{\citenamefont {Nanda}\ and\ \citenamefont
  {Krylov}(2016)}]{Nanda2016}%
  \BibitemOpen
  \bibfield  {author} {\bibinfo {author} {\bibfnamefont {K.~D.}\ \bibnamefont
  {Nanda}}\ and\ \bibinfo {author} {\bibfnamefont {A.~I.}\ \bibnamefont
  {Krylov}},\ }\bibfield  {title} {\bibinfo {title} {{Static polarizabilities
  for excited states within the spin-conserving and spin-flipping
  equation-of-motion coupled-cluster singles and doubles formalism: Theory,
  implementation, and benchmarks}},\ }\href {https://doi.org/10.1063/1.4967860}
  {\bibfield  {journal} {\bibinfo  {journal} {The Journal of Chemical Physics}\
  }\textbf {\bibinfo {volume} {145}},\ \bibinfo {pages} {204116} (\bibinfo
  {year} {2016})}\BibitemShut {NoStop}%
\bibitem [{\citenamefont {Epifanovsky}\ \emph {et~al.}(2021)\citenamefont
  {Epifanovsky}, \citenamefont {Gilbert}, \citenamefont {Feng}, \citenamefont
  {Lee}, \citenamefont {Mao}, \citenamefont {Mardirossian}, \citenamefont
  {Pokhilko}, \citenamefont {White}, \citenamefont {Coons}, \citenamefont
  {Dempwolff}, \citenamefont {Gan}, \citenamefont {Hait}, \citenamefont {Horn},
  \citenamefont {Jacobson}, \citenamefont {Kaliman} \emph {et~al.}}]{QChem5}%
  \BibitemOpen
  \bibfield  {author} {\bibinfo {author} {\bibfnamefont {E.}~\bibnamefont
  {Epifanovsky}}, \bibinfo {author} {\bibfnamefont {A.~T.~B.}\ \bibnamefont
  {Gilbert}}, \bibinfo {author} {\bibfnamefont {X.}~\bibnamefont {Feng}},
  \bibinfo {author} {\bibfnamefont {J.}~\bibnamefont {Lee}}, \bibinfo {author}
  {\bibfnamefont {Y.}~\bibnamefont {Mao}}, \bibinfo {author} {\bibfnamefont
  {N.}~\bibnamefont {Mardirossian}}, \bibinfo {author} {\bibfnamefont
  {P.}~\bibnamefont {Pokhilko}}, \bibinfo {author} {\bibfnamefont {A.~F.}\
  \bibnamefont {White}}, \bibinfo {author} {\bibfnamefont {M.~P.}\ \bibnamefont
  {Coons}}, \bibinfo {author} {\bibfnamefont {A.~L.}\ \bibnamefont
  {Dempwolff}}, \bibinfo {author} {\bibfnamefont {Z.}~\bibnamefont {Gan}},
  \bibinfo {author} {\bibfnamefont {D.}~\bibnamefont {Hait}}, \bibinfo {author}
  {\bibfnamefont {P.~R.}\ \bibnamefont {Horn}}, \bibinfo {author}
  {\bibfnamefont {L.~D.}\ \bibnamefont {Jacobson}}, \bibinfo {author}
  {\bibfnamefont {I.}~\bibnamefont {Kaliman}}, \emph {et~al.},\ }\bibfield
  {title} {\bibinfo {title} {{Software for the frontiers of quantum chemistry:
  An overview of developments in the Q-Chem 5 package}},\ }\href
  {https://doi.org/10.1063/5.0055522} {\bibfield  {journal} {\bibinfo
  {journal} {The Journal of Chemical Physics}\ }\textbf {\bibinfo {volume}
  {155}},\ \bibinfo {pages} {084801} (\bibinfo {year} {2021})}\BibitemShut
  {NoStop}%
\bibitem [{\citenamefont {Lim}\ \emph {et~al.}(2006)\citenamefont {Lim},
  \citenamefont {Stoll},\ and\ \citenamefont {Schwerdtfeger}}]{Lim2006}%
  \BibitemOpen
  \bibfield  {author} {\bibinfo {author} {\bibfnamefont {I.~S.}\ \bibnamefont
  {Lim}}, \bibinfo {author} {\bibfnamefont {H.}~\bibnamefont {Stoll}},\ and\
  \bibinfo {author} {\bibfnamefont {P.}~\bibnamefont {Schwerdtfeger}},\
  }\bibfield  {title} {\bibinfo {title} {{Relativistic small-core
  energy-consistent pseudopotentials for the alkaline-earth elements from Ca to
  Ra}},\ }\href {https://doi.org/10.1063/1.2148945} {\bibfield  {journal}
  {\bibinfo  {journal} {The Journal of Chemical Physics}\ }\textbf {\bibinfo
  {volume} {124}},\ \bibinfo {pages} {034107} (\bibinfo {year}
  {2006})}\BibitemShut {NoStop}%
\bibitem [{\citenamefont {Dalgarno}\ \emph {et~al.}(1971)\citenamefont
  {Dalgarno}, \citenamefont {Ford},\ and\ \citenamefont
  {Browne}}]{Dalgarno1971}%
  \BibitemOpen
  \bibfield  {author} {\bibinfo {author} {\bibfnamefont {A.}~\bibnamefont
  {Dalgarno}}, \bibinfo {author} {\bibfnamefont {A.~L.}\ \bibnamefont {Ford}},\
  and\ \bibinfo {author} {\bibfnamefont {J.~C.}\ \bibnamefont {Browne}},\
  }\bibfield  {title} {\bibinfo {title} {{Direct Sum-of-States Calculations of
  the Frequency-Dependent Polarizability of H$_2$}},\ }\href
  {https://doi.org/10.1103/PhysRevLett.27.1033} {\bibfield  {journal} {\bibinfo
   {journal} {Physical Review Letters}\ }\textbf {\bibinfo {volume} {27}},\
  \bibinfo {pages} {1033} (\bibinfo {year} {1971})}\BibitemShut {NoStop}%
\bibitem [{\citenamefont {Brown}\ and\ \citenamefont
  {Carrington}(2003)}]{Brown2003}%
  \BibitemOpen
  \bibfield  {author} {\bibinfo {author} {\bibfnamefont {J.~M.}\ \bibnamefont
  {Brown}}\ and\ \bibinfo {author} {\bibfnamefont {A.}~\bibnamefont
  {Carrington}},\ }\href {https://doi.org/10.1017/CBO9780511814808} {\emph
  {\bibinfo {title} {{Rotational Spectroscopy of Diatomic Molecules}}}}\
  (\bibinfo  {publisher} {Cambridge University Press},\ \bibinfo {address}
  {Cambridge},\ \bibinfo {year} {2003})\BibitemShut {NoStop}%
\bibitem [{\citenamefont {Heijmen}\ \emph {et~al.}(1996)\citenamefont
  {Heijmen}, \citenamefont {Moszynski}, \citenamefont {Wormer},\ and\
  \citenamefont {{van der Avoird}}}]{Heijmen1996}%
  \BibitemOpen
  \bibfield  {author} {\bibinfo {author} {\bibfnamefont {T.~G.~A.}\
  \bibnamefont {Heijmen}}, \bibinfo {author} {\bibfnamefont {R.}~\bibnamefont
  {Moszynski}}, \bibinfo {author} {\bibfnamefont {P.~E.~S.}\ \bibnamefont
  {Wormer}},\ and\ \bibinfo {author} {\bibfnamefont {A.}~\bibnamefont {{van der
  Avoird}}},\ }\bibfield  {title} {\bibinfo {title} {{Symmetry-adapted
  perturbation theory applied to interaction-induced properties of collisional
  complexes}},\ }\href {https://doi.org/10.1080/002689796174029} {\bibfield
  {journal} {\bibinfo  {journal} {Molecular Physics}\ }\textbf {\bibinfo
  {volume} {89}},\ \bibinfo {pages} {81} (\bibinfo {year} {1996})}\BibitemShut
  {NoStop}%
\bibitem [{\citenamefont {Colbert}\ and\ \citenamefont
  {Miller}(1992)}]{Colbert1992}%
  \BibitemOpen
  \bibfield  {author} {\bibinfo {author} {\bibfnamefont {D.~T.}\ \bibnamefont
  {Colbert}}\ and\ \bibinfo {author} {\bibfnamefont {W.~H.}\ \bibnamefont
  {Miller}},\ }\bibfield  {title} {\bibinfo {title} {{A novel discrete variable
  representation for quantum mechanical reactive scattering via the S-matrix
  Kohn method}},\ }\href {https://doi.org/10.1063/1.462100} {\bibfield
  {journal} {\bibinfo  {journal} {J. Chem. Phys.}\ }\textbf {\bibinfo {volume}
  {96}},\ \bibinfo {pages} {1982} (\bibinfo {year} {1992})}\BibitemShut
  {NoStop}%
\bibitem [{\citenamefont {Tiesinga}\ \emph {et~al.}(1998)\citenamefont
  {Tiesinga}, \citenamefont {Williams},\ and\ \citenamefont
  {Julienne}}]{Tiesinga1998}%
  \BibitemOpen
  \bibfield  {author} {\bibinfo {author} {\bibfnamefont {E.}~\bibnamefont
  {Tiesinga}}, \bibinfo {author} {\bibfnamefont {C.~J.}\ \bibnamefont
  {Williams}},\ and\ \bibinfo {author} {\bibfnamefont {P.~S.}\ \bibnamefont
  {Julienne}},\ }\bibfield  {title} {\bibinfo {title} {{Photoassociative
  spectroscopy of highly excited vibrational levels of alkali-metal dimers:
  Green-function approach for eigenvalue solvers}},\ }\href
  {https://doi.org/10.1103/PhysRevA.57.4257} {\bibfield  {journal} {\bibinfo
  {journal} {Phys. Rev. A}\ }\textbf {\bibinfo {volume} {57}},\ \bibinfo
  {pages} {4257} (\bibinfo {year} {1998})}\BibitemShut {NoStop}%
\bibitem [{\citenamefont {Stein}\ \emph {et~al.}(2010)\citenamefont {Stein},
  \citenamefont {Kn{\"{o}}ckel},\ and\ \citenamefont {Tiemann}}]{Stein2010}%
  \BibitemOpen
  \bibfield  {author} {\bibinfo {author} {\bibfnamefont {A.}~\bibnamefont
  {Stein}}, \bibinfo {author} {\bibfnamefont {H.}~\bibnamefont
  {Kn{\"{o}}ckel}},\ and\ \bibinfo {author} {\bibfnamefont {E.}~\bibnamefont
  {Tiemann}},\ }\bibfield  {title} {\bibinfo {title} {{{$^1$S+$^1$S asymptote
  of Sr$_2$ studied by Fourier-transform spectroscopy}}},\ }\href
  {https://doi.org/10.1140/epjd/e2010-00058-y} {\bibfield  {journal} {\bibinfo
  {journal} {Eur. Phys. J. D}\ }\textbf {\bibinfo {volume} {57}},\ \bibinfo
  {pages} {171} (\bibinfo {year} {2010})}\BibitemShut {NoStop}%
\bibitem [{\citenamefont {Fraser}(1954)}]{Fraser1954}%
  \BibitemOpen
  \bibfield  {author} {\bibinfo {author} {\bibfnamefont {P.~A.}\ \bibnamefont
  {Fraser}},\ }\bibfield  {title} {\bibinfo {title} {A method of determining
  the electronic transition moment for diatomic molecules},\ }\href
  {https://doi.org/10.1139/p54-054} {\bibfield  {journal} {\bibinfo  {journal}
  {Canadian Journal of Physics}\ }\textbf {\bibinfo {volume} {32}},\ \bibinfo
  {pages} {515} (\bibinfo {year} {1954})}\BibitemShut {NoStop}%
\bibitem [{\citenamefont {Clementi}\ \emph {et~al.}(1967)\citenamefont
  {Clementi}, \citenamefont {Raimondi},\ and\ \citenamefont
  {Reinhardt}}]{Clementi1967}%
  \BibitemOpen
  \bibfield  {author} {\bibinfo {author} {\bibfnamefont {E.}~\bibnamefont
  {Clementi}}, \bibinfo {author} {\bibfnamefont {D.~L.}\ \bibnamefont
  {Raimondi}},\ and\ \bibinfo {author} {\bibfnamefont {W.~P.}\ \bibnamefont
  {Reinhardt}},\ }\bibfield  {title} {\bibinfo {title} {{Atomic Screening
  Constants from SCF Functions. II. Atoms with 37 to 86 Electrons}},\ }\href
  {https://doi.org/10.1063/1.1712084} {\bibfield  {journal} {\bibinfo
  {journal} {J. Chem. Phys.}\ }\textbf {\bibinfo {volume} {471}},\ \bibinfo
  {pages} {1300} (\bibinfo {year} {1967})}\BibitemShut {NoStop}%
\bibitem [{\citenamefont {Taylor}\ and\ \citenamefont
  {Kuyatt}(1994)}]{Taylor1994}%
  \BibitemOpen
  \bibfield  {author} {\bibinfo {author} {\bibfnamefont {B.~N.}\ \bibnamefont
  {Taylor}}\ and\ \bibinfo {author} {\bibfnamefont {C.~E.}\ \bibnamefont
  {Kuyatt}},\ }\href {https://doi.org/10.6028/NIST.TN.1297} {\emph {\bibinfo
  {title} {NIST Technical Note 1297. Guidelines for evaluating and expressing
  the uncertainty of NIST measurement results}}}\ (\bibinfo  {publisher} {US
  Department of Commerce, Technology Administration, National Institute of
  Standards and Technology},\ \bibinfo {year} {1994})\BibitemShut {NoStop}%
\bibitem [{\citenamefont {Farley}\ and\ \citenamefont
  {Wing}(1981)}]{Farley1981}%
  \BibitemOpen
  \bibfield  {author} {\bibinfo {author} {\bibfnamefont {J.~W.}\ \bibnamefont
  {Farley}}\ and\ \bibinfo {author} {\bibfnamefont {W.~H.}\ \bibnamefont
  {Wing}},\ }\bibfield  {title} {\bibinfo {title} {Accurate calculation of
  dynamic stark shifts and depopulation rates of {R}ydberg energy levels
  induced by blackbody radiation. {H}ydrogen, helium, and alkali-metal atoms},\
  }\href {https://doi.org/10.1103/PhysRevA.23.2397} {\bibfield  {journal}
  {\bibinfo  {journal} {Phys. Rev. A}\ }\textbf {\bibinfo {volume} {23}},\
  \bibinfo {pages} {2397} (\bibinfo {year} {1981})}\BibitemShut {NoStop}%
\bibitem [{\citenamefont {Mitroy}\ \emph {et~al.}(2010)\citenamefont {Mitroy},
  \citenamefont {Safronova},\ and\ \citenamefont {Clark}}]{Mitroy2010}%
  \BibitemOpen
  \bibfield  {author} {\bibinfo {author} {\bibfnamefont {J.}~\bibnamefont
  {Mitroy}}, \bibinfo {author} {\bibfnamefont {M.~S.}\ \bibnamefont
  {Safronova}},\ and\ \bibinfo {author} {\bibfnamefont {C.~W.}\ \bibnamefont
  {Clark}},\ }\bibfield  {title} {\bibinfo {title} {{Theory and applications of
  atomic and ionic polarizabilities}},\ }\href
  {https://doi.org/10.1088/0953-4075/43/20/202001} {\bibfield  {journal}
  {\bibinfo  {journal} {Journal of Physics B: Atomic, Molecular and Optical
  Physics}\ }\textbf {\bibinfo {volume} {43}},\ \bibinfo {pages} {202001}
  (\bibinfo {year} {2010})}\BibitemShut {NoStop}%
\bibitem [{\citenamefont {Bergmann}\ \emph {et~al.}(1998)\citenamefont
  {Bergmann}, \citenamefont {Theuer},\ and\ \citenamefont
  {Shore}}]{Bergmann1998}%
  \BibitemOpen
  \bibfield  {author} {\bibinfo {author} {\bibfnamefont {K.}~\bibnamefont
  {Bergmann}}, \bibinfo {author} {\bibfnamefont {H.}~\bibnamefont {Theuer}},\
  and\ \bibinfo {author} {\bibfnamefont {B.~W.}\ \bibnamefont {Shore}},\
  }\bibfield  {title} {\bibinfo {title} {{Coherent population transfer among
  quantum states of atoms and molecules}},\ }\href
  {https://doi.org/10.1103/RevModPhys.70.1003} {\bibfield  {journal} {\bibinfo
  {journal} {Reviews of Modern Physics}\ }\textbf {\bibinfo {volume} {70}},\
  \bibinfo {pages} {1003} (\bibinfo {year} {1998})}\BibitemShut {NoStop}%
\bibitem [{\citenamefont {Vitanov}\ \emph {et~al.}(2017)\citenamefont
  {Vitanov}, \citenamefont {Rangelov}, \citenamefont {Shore},\ and\
  \citenamefont {Bergmann}}]{Vitanov2017}%
  \BibitemOpen
  \bibfield  {author} {\bibinfo {author} {\bibfnamefont {N.~V.}\ \bibnamefont
  {Vitanov}}, \bibinfo {author} {\bibfnamefont {A.~A.}\ \bibnamefont
  {Rangelov}}, \bibinfo {author} {\bibfnamefont {B.~W.}\ \bibnamefont
  {Shore}},\ and\ \bibinfo {author} {\bibfnamefont {K.}~\bibnamefont
  {Bergmann}},\ }\bibfield  {title} {\bibinfo {title} {{Stimulated Raman
  adiabatic passage in physics, chemistry, and beyond}},\ }\href
  {https://doi.org/10.1103/RevModPhys.89.015006} {\bibfield  {journal}
  {\bibinfo  {journal} {Reviews of Modern Physics}\ }\textbf {\bibinfo {volume}
  {89}},\ \bibinfo {pages} {1} (\bibinfo {year} {2017})}\BibitemShut {NoStop}%
\bibitem [{\citenamefont {Chen}(2006)}]{JingbiaoChen2006}%
  \BibitemOpen
  \bibfield  {author} {\bibinfo {author} {\bibfnamefont {J.}~\bibnamefont
  {Chen}},\ }\bibfield  {title} {\bibinfo {title} {{To simulate blackbody
  radiation frequency shift in cesium fountain frequency standard with CO$_2$
  laser}},\ }\href {https://doi.org/10.1109/tuffc.2006.1678197} {\bibfield
  {journal} {\bibinfo  {journal} {{IEEE} Transactions on Ultrasonics,
  Ferroelectrics and Frequency Control}\ }\textbf {\bibinfo {volume} {53}},\
  \bibinfo {pages} {1685} (\bibinfo {year} {2006})}\BibitemShut {NoStop}%
\bibitem [{\citenamefont {Arnold}\ \emph {et~al.}(2018)\citenamefont {Arnold},
  \citenamefont {Kaewuam}, \citenamefont {Roy}, \citenamefont {Tan},\ and\
  \citenamefont {Barrett}}]{Arnold2018}%
  \BibitemOpen
  \bibfield  {author} {\bibinfo {author} {\bibfnamefont {K.~J.}\ \bibnamefont
  {Arnold}}, \bibinfo {author} {\bibfnamefont {R.}~\bibnamefont {Kaewuam}},
  \bibinfo {author} {\bibfnamefont {A.}~\bibnamefont {Roy}}, \bibinfo {author}
  {\bibfnamefont {T.~R.}\ \bibnamefont {Tan}},\ and\ \bibinfo {author}
  {\bibfnamefont {M.~D.}\ \bibnamefont {Barrett}},\ }\bibfield  {title}
  {\bibinfo {title} {Blackbody radiation shift assessment for a lutetium ion
  clock},\ }\href {https://doi.org/10.1038/s41467-018-04079-x} {\bibfield
  {journal} {\bibinfo  {journal} {Nature Communications}\ }\textbf {\bibinfo
  {volume} {9}},\ \bibinfo {pages} {1650} (\bibinfo {year} {2018})}\BibitemShut
  {NoStop}%
\end{thebibliography}%


\begin{thebibliography}{8}%
\makeatletter
\providecommand \@ifxundefined [1]{%
 \@ifx{#1\undefined}
}%
\providecommand \@ifnum [1]{%
 \ifnum #1\expandafter \@firstoftwo
 \else \expandafter \@secondoftwo
 \fi
}%
\providecommand \@ifx [1]{%
 \ifx #1\expandafter \@firstoftwo
 \else \expandafter \@secondoftwo
 \fi
}%
\providecommand \natexlab [1]{#1}%
\providecommand \enquote  [1]{``#1''}%
\providecommand \bibnamefont  [1]{#1}%
\providecommand \bibfnamefont [1]{#1}%
\providecommand \citenamefont [1]{#1}%
\providecommand \href@noop [0]{\@secondoftwo}%
\providecommand \href [0]{\begingroup \@sanitize@url \@href}%
\providecommand \@href[1]{\@@startlink{#1}\@@href}%
\providecommand \@@href[1]{\endgroup#1\@@endlink}%
\providecommand \@sanitize@url [0]{\catcode `\\12\catcode `\$12\catcode
  `\&12\catcode `\#12\catcode `\^12\catcode `\_12\catcode `\%12\relax}%
\providecommand \@@startlink[1]{}%
\providecommand \@@endlink[0]{}%
\providecommand \url  [0]{\begingroup\@sanitize@url \@url }%
\providecommand \@url [1]{\endgroup\@href {#1}{\urlprefix }}%
\providecommand \urlprefix  [0]{URL }%
\providecommand \Eprint [0]{\href }%
\providecommand \doibase [0]{https://doi.org/}%
\providecommand \selectlanguage [0]{\@gobble}%
\providecommand \bibinfo  [0]{\@secondoftwo}%
\providecommand \bibfield  [0]{\@secondoftwo}%
\providecommand \translation [1]{[#1]}%
\providecommand \BibitemOpen [0]{}%
\providecommand \bibitemStop [0]{}%
\providecommand \bibitemNoStop [0]{.\EOS\space}%
\providecommand \EOS [0]{\spacefactor3000\relax}%
\providecommand \BibitemShut  [1]{\csname bibitem#1\endcsname}%
\let\auto@bib@innerbib\@empty
\bibitem [{\citenamefont {Leung}\ \emph {et~al.}(2020)\citenamefont {Leung},
  \citenamefont {Majewska}, \citenamefont {Bekker}, \citenamefont {Lee},
  \citenamefont {Tiberi}, \citenamefont {Kondov}, \citenamefont {Moszynski},\
  and\ \citenamefont {Zelevinsky}}]{Leung2020}%
  \BibitemOpen
  \bibfield  {author} {\bibinfo {author} {\bibfnamefont {K.~H.}\ \bibnamefont
  {Leung}}, \bibinfo {author} {\bibfnamefont {I.}~\bibnamefont {Majewska}},
  \bibinfo {author} {\bibfnamefont {H.}~\bibnamefont {Bekker}}, \bibinfo
  {author} {\bibfnamefont {C.-H.}\ \bibnamefont {Lee}}, \bibinfo {author}
  {\bibfnamefont {E.}~\bibnamefont {Tiberi}}, \bibinfo {author} {\bibfnamefont
  {S.~S.}\ \bibnamefont {Kondov}}, \bibinfo {author} {\bibfnamefont
  {R.}~\bibnamefont {Moszynski}},\ and\ \bibinfo {author} {\bibfnamefont
  {T.}~\bibnamefont {Zelevinsky}},\ }\bibfield  {title} {\bibinfo {title}
  {Transition strength measurements to guide magic wavelength selection in
  optically trapped molecules},\ }\href
  {https://doi.org/10.1103/PhysRevLett.125.153001} {\bibfield  {journal}
  {\bibinfo  {journal} {Phys. Rev. Lett.}\ }\textbf {\bibinfo {volume} {125}},\
  \bibinfo {pages} {153001} (\bibinfo {year} {2020})}\BibitemShut {NoStop}%
\bibitem [{\citenamefont {Leung}\ \emph {et~al.}(2023)\citenamefont {Leung},
  \citenamefont {Iritani}, \citenamefont {Tiberi}, \citenamefont {Majewska},
  \citenamefont {Borkowski}, \citenamefont {Moszynski},\ and\ \citenamefont
  {Zelevinsky}}]{Leung2023}%
  \BibitemOpen
  \bibfield  {author} {\bibinfo {author} {\bibfnamefont {K.~H.}\ \bibnamefont
  {Leung}}, \bibinfo {author} {\bibfnamefont {B.}~\bibnamefont {Iritani}},
  \bibinfo {author} {\bibfnamefont {E.}~\bibnamefont {Tiberi}}, \bibinfo
  {author} {\bibfnamefont {I.}~\bibnamefont {Majewska}}, \bibinfo {author}
  {\bibfnamefont {M.}~\bibnamefont {Borkowski}}, \bibinfo {author}
  {\bibfnamefont {R.}~\bibnamefont {Moszynski}},\ and\ \bibinfo {author}
  {\bibfnamefont {T.}~\bibnamefont {Zelevinsky}},\ }\bibfield  {title}
  {\bibinfo {title} {Terahertz vibrational molecular clock with systematic
  uncertainty at the ${10}^{\ensuremath{-}14}$ level},\ }\href
  {https://doi.org/10.1103/PhysRevX.13.011047} {\bibfield  {journal} {\bibinfo
  {journal} {Phys. Rev. X}\ }\textbf {\bibinfo {volume} {13}},\ \bibinfo
  {pages} {011047} (\bibinfo {year} {2023})}\BibitemShut {NoStop}%
\bibitem [{\citenamefont {Safronova}(2023)}]{SafronovaPrivateCommunication}%
  \BibitemOpen
  \bibfield  {author} {\bibinfo {author} {\bibfnamefont {M.~S.}\ \bibnamefont
  {Safronova}},\ }\href@noop {} {}\bibinfo {howpublished} {{private
  communication}} (\bibinfo {year} {2023})\BibitemShut {NoStop}%
\bibitem [{\citenamefont {Bonin}\ and\ \citenamefont
  {Kresin}(1997)}]{Bonin1997}%
  \BibitemOpen
  \bibfield  {author} {\bibinfo {author} {\bibfnamefont {K.~D.}\ \bibnamefont
  {Bonin}}\ and\ \bibinfo {author} {\bibfnamefont {V.~V.}\ \bibnamefont
  {Kresin}},\ }\href@noop {} {\emph {\bibinfo {title} {{Electric-dipole
  polarizabilities of atoms, molecules, and clusters}}}}\ (\bibinfo
  {publisher} {World Scientific},\ \bibinfo {year} {1997})\BibitemShut
  {NoStop}%
\bibitem [{\citenamefont {Kondov}\ \emph {et~al.}(2019)\citenamefont {Kondov},
  \citenamefont {Lee}, \citenamefont {Leung}, \citenamefont {Liedl},
  \citenamefont {Majewska}, \citenamefont {Moszynski},\ and\ \citenamefont
  {Zelevinsky}}]{Kondov2019a}%
  \BibitemOpen
  \bibfield  {author} {\bibinfo {author} {\bibfnamefont {S.~S.}\ \bibnamefont
  {Kondov}}, \bibinfo {author} {\bibfnamefont {C.-H.}\ \bibnamefont {Lee}},
  \bibinfo {author} {\bibfnamefont {K.~H.}\ \bibnamefont {Leung}}, \bibinfo
  {author} {\bibfnamefont {C.}~\bibnamefont {Liedl}}, \bibinfo {author}
  {\bibfnamefont {I.}~\bibnamefont {Majewska}}, \bibinfo {author}
  {\bibfnamefont {R.}~\bibnamefont {Moszynski}},\ and\ \bibinfo {author}
  {\bibfnamefont {T.}~\bibnamefont {Zelevinsky}},\ }\bibfield  {title}
  {\bibinfo {title} {Molecular lattice clock with long vibrational coherence},\
  }\href {https://doi.org/10.1038/s41567-019-0632-3} {\bibfield  {journal}
  {\bibinfo  {journal} {Nature Physics}\ }\textbf {\bibinfo {volume} {15}},\
  \bibinfo {pages} {1118} (\bibinfo {year} {2019})}\BibitemShut {NoStop}%
\bibitem [{\citenamefont {Landau}\ and\ \citenamefont
  {Lifshitz}(1958)}]{LandauStatisticalPhysics}%
  \BibitemOpen
  \bibfield  {author} {\bibinfo {author} {\bibfnamefont {L.}~\bibnamefont
  {Landau}}\ and\ \bibinfo {author} {\bibfnamefont {E.}~\bibnamefont
  {Lifshitz}},\ }\href@noop {} {\emph {\bibinfo {title} {Statistical
  Physics}}}\ (\bibinfo  {publisher} {Pergamon Press},\ \bibinfo {year}
  {1958})\BibitemShut {NoStop}%
\bibitem [{\citenamefont {Stein}\ \emph {et~al.}(2010)\citenamefont {Stein},
  \citenamefont {Kn{\"{o}}ckel},\ and\ \citenamefont {Tiemann}}]{Stein2010}%
  \BibitemOpen
  \bibfield  {author} {\bibinfo {author} {\bibfnamefont {A.}~\bibnamefont
  {Stein}}, \bibinfo {author} {\bibfnamefont {H.}~\bibnamefont
  {Kn{\"{o}}ckel}},\ and\ \bibinfo {author} {\bibfnamefont {E.}~\bibnamefont
  {Tiemann}},\ }\bibfield  {title} {\bibinfo {title} {{{$^1$S+$^1$S asymptote
  of Sr$_2$ studied by Fourier-transform spectroscopy}}},\ }\href
  {https://doi.org/10.1140/epjd/e2010-00058-y} {\bibfield  {journal} {\bibinfo
  {journal} {Eur. Phys. J. D}\ }\textbf {\bibinfo {volume} {57}},\ \bibinfo
  {pages} {171} (\bibinfo {year} {2010})}\BibitemShut {NoStop}%
\bibitem [{\citenamefont {{Le Roy}}\ \emph {et~al.}(2009)\citenamefont {{Le
  Roy}}, \citenamefont {Dattani}, \citenamefont {Coxon}, \citenamefont {Ross},
  \citenamefont {Crozet},\ and\ \citenamefont {Linton}}]{LeRoy2009}%
  \BibitemOpen
  \bibfield  {author} {\bibinfo {author} {\bibfnamefont {R.~J.}\ \bibnamefont
  {{Le Roy}}}, \bibinfo {author} {\bibfnamefont {N.~S.}\ \bibnamefont
  {Dattani}}, \bibinfo {author} {\bibfnamefont {J.~A.}\ \bibnamefont {Coxon}},
  \bibinfo {author} {\bibfnamefont {A.~J.}\ \bibnamefont {Ross}}, \bibinfo
  {author} {\bibfnamefont {P.}~\bibnamefont {Crozet}},\ and\ \bibinfo {author}
  {\bibfnamefont {C.}~\bibnamefont {Linton}},\ }\bibfield  {title} {\bibinfo
  {title} {{Accurate analytic potentials for Li$_2$(X $^1$$\Sigma$$_g$) and
  Li$_2$(A X $^1$$\Sigma$$_u^+$) from 2 to 90 {\AA}, and the radiative lifetime
  of Li(2p)}},\ }\href {https://doi.org/10.1063/1.3264688} {\bibfield
  {journal} {\bibinfo  {journal} {The Journal of Chemical Physics}\ }\textbf
  {\bibinfo {volume} {131}},\ \bibinfo {pages} {204309} (\bibinfo {year}
  {2009})}\BibitemShut {NoStop}%
\end{thebibliography}%

\end{document}


\renewcommand{\theequation}{S\arabic{equation}}
\renewcommand{\thefigure}{S\arabic{figure}}
\renewcommand{\bibnumfmt}[1]{[S#1]}
\renewcommand{\citenumfont}[1]{S#1}
\renewcommand{\thetable}{S\arabic{table}}

\title{Supplemental Material: Accurate Determination of Blackbody Radiation Shifts in a Strontium Molecular Lattice Clock}

    \author{B. Iritani\orcidlink{0000-0002-7911-2755}}
        \altaffiliation{These authors contributed equally to this work.}
        \affiliation{Department of Physics, Columbia University, 538 West 120th Street, New York, NY 10027-5255, USA}
        
    \author{E. Tiberi\orcidlink{0000-0001-7168-7194}}
        \altaffiliation{These authors contributed equally to this work.}
        \affiliation{Department of Physics, Columbia University, 538 West 120th Street, New York, NY 10027-5255, USA}

    \author{W. Skomorowski\orcidlink{0000-0002-0364-435X}}
    \affiliation{Centre of New Technologies, University of Warsaw, Banacha 2c, 02-097 Warsaw, Poland}

    \author{R. Moszynski\orcidlink{0009-0008-7669-3751}}
        \affiliation{Quantum Chemistry Laboratory, Department of Chemistry,
        University of Warsaw, Pasteura 1, 02-093 Warsaw, Poland}
        
     \author{M. Borkowski\orcidlink{0000-0003-0236-8100}}
        \email{mateusz@cold-molecules.com}   
        \affiliation{Department of Physics, Columbia University, 538 West 120th Street, New York, NY 10027-5255, USA}
        \affiliation{Van der Waals-Zeeman Institute, Institute of Physics, University of Amsterdam, Science Park 904, 1098 XH Amsterdam, The Netherlands}
        \affiliation{Institute of Physics, Faculty of Physics, Astronomy and Informatics, Nicolaus Copernicus University, Grudziadzka 5, 87-100 Torun, Poland}
        
    \author{T. Zelevinsky\orcidlink{0000-0003-3682-4901}}
        \email{tanya.zelevinsky@columbia.edu}
        \affiliation{Department of Physics, Columbia University, 538 West 120th Street, New York, NY 10027-5255, USA}

\maketitle

\section*{S1. Determination of magic wavelengths}

\begin{figure}[b]
    \includegraphics[width=\columnwidth]{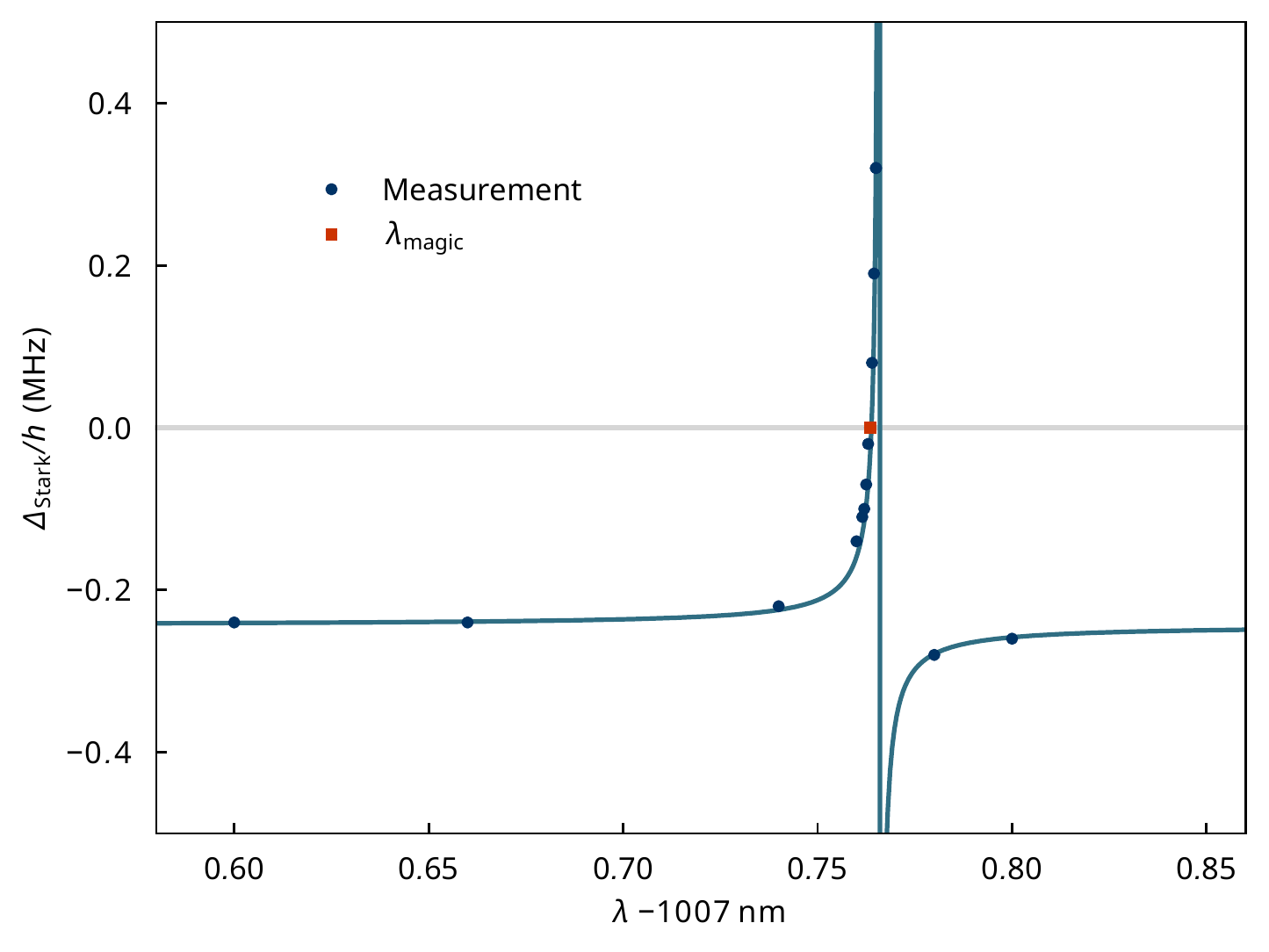}
    \caption{Search for the magic wavelength on the example of the $v = 12\,\leftrightarrow\,v' = 62$ molecular clock transition. Points denote the experimental lattice-induced ac Stark shift as a function of lattice laser wavelength $\lambda$. The fitted function is Eq.~(\ref{eq:magic}). The red square indicates the magic wavelength $\lambda_{\rm magic} = 1007.7634(10)\,{\rm nm}$, where the Stark shift $\Delta_{\rm Stark}$ is cancelled out.\label{fig:magic}}
\end{figure}

For each of the investigated molecular clock transitions we have determined its corresponding magic wavelength. Our process for finding magic wavelengths consists of several steps and combines theoretical modelling and experiment.

The polarizability of the initial weakly-bound molecular state is approximately twice the polarizability of the constituent strontium atoms and only has a very weak dependence on the wavelength of the lattice laser. On the other hand, the polarizability for the deeply-bound states has many resonances due to strong transitions to the vibrational states supported by the $1_u$ state correlating to the $^1$S$_0$+$^3$P$_1$ asymptote. We exploit this to tune the polarizability of the deeply-bound state to that of the weakly-bound state to achieve the magic condition.

We first employ a theoretical interaction model to calculate transition dipole moments for transitions from the deeply bound  molecular clock state $v$ to vibrational states in the $(1)\,1_u$ excited-state potential. We select $(1)\,1_u$ states such that the line strength $S$ is greater than $\sim$$10^{-5}\,e^2 a_0^2$ (here $e$ is the electron charge, $a_0$ is the Bohr radius)~\cite{Leung2020}. Then, we predict the magic wavelengths by calculating the differential polarizability of the clock transition using a sum-over-states approach. This provides a starting point for the final experimental search. By varying the power of the lattice beam, we measure the Stark shift $\Delta_{\rm Stark}$ of the molecular clock line at several wavelengths spread over $\sim$10 GHz centered around the predicted magic wavelength (Figure~\ref{fig:magic}). Then, we fit a simple formula
        \begin{equation}
            \Delta_{\rm Stark}(\lambda) = \frac{a}{\lambda-\lambda_0} + b \label{eq:magic}
        \end{equation}
        to lattice Stark shifts measured as a function of frequency to find the zero crossing, $\lambda_{\rm magic}=\lambda_0 - a/b$. Our lattice wavelength is stabilized to a wavemeter at $\sim$30\,MHz precision. It should also be pointed out that the absolute calibration of the wavemeter is on the order of $0.001\,{\rm nm}$ as indicated in Table~1 in the main text.

        \section*{S2. Measurements of line positions}

        We measure the relative binding energies by scanning the frequency difference between the two Raman lasers, detuned by +30 MHz from the intermediate state, a molecular level of $(1)\,0_u^+$ symmetry. The Raman pump laser is locked to a high finesse cavity, and the repetition rate of an optical frequency comb is in turn referenced to this laser. The carrier envelope offset of the frequency comb, as well as acousto-optical modulators used on both Raman lasers, are referenced to a commercial rubidium clock at a $\sim$$10^{-12}$ precision \cite{Leung2023}. Finally, the anti-Stokes laser of the Raman pair is phase-locked to the optical frequency comb.

\begin{figure}
        \includegraphics[width=\columnwidth]{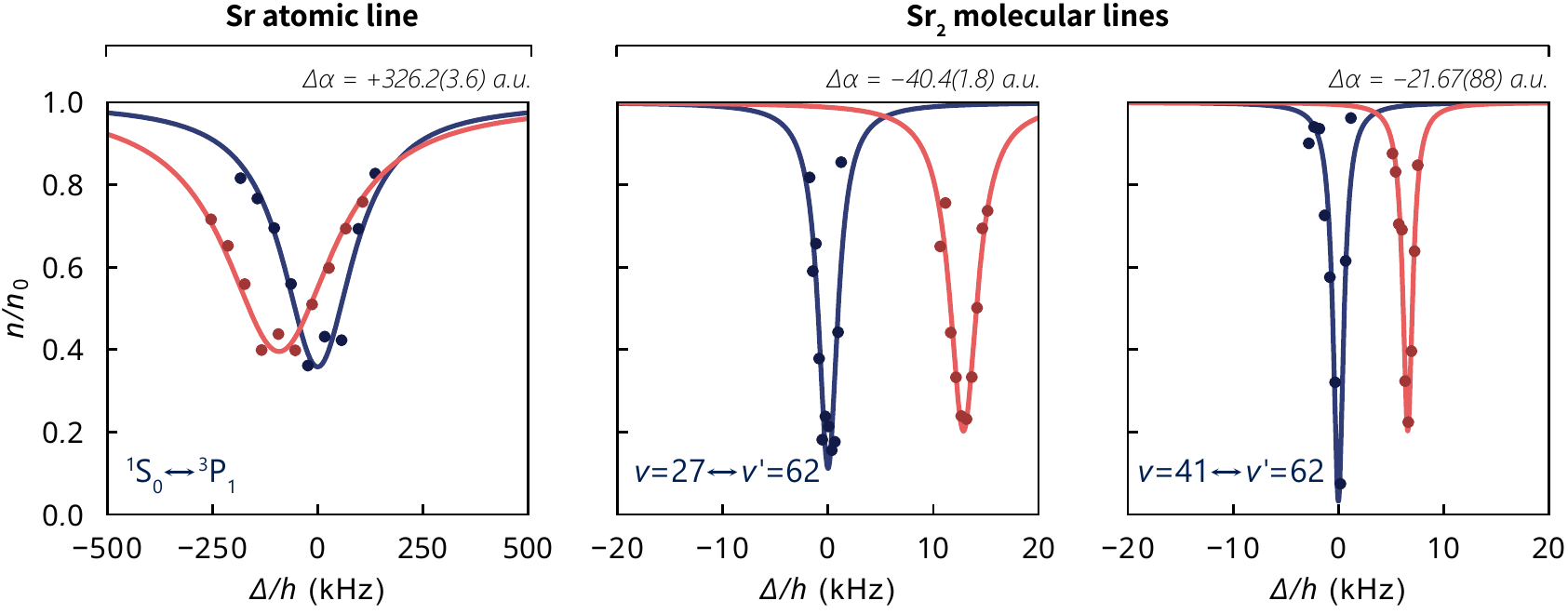}
        \caption{Example lineshapes seen in our ac Stark shift measurements. To determine the differential ac polarizability, we measure a lineshape each with the extra ac Stark laser off (dark blue) and on (light red). To determine the differential polarizabilities $\Delta\alpha$ we measure ratios of ac Stark shift slopes between different transitions. As an absolute reference we used the narrow $^1$S$_0$$\leftrightarrow$$^3$P$_1$ atomic intercombination transition with a known $\Delta\alpha = +326.2(3.6)$~a.u.~\cite{SafronovaPrivateCommunication}. The molecular ac Stark shifts would be compared to a common $27\leftrightarrow62$ transition which would then be calibrated to the atomic line. }
        \label{fig:atomicCalibration}
    \end{figure}

    The line positions are measured through scans of the relative Raman frequency (Figure~\ref{fig:atomicCalibration}). We fit the scans with a Lorentzian lineshape with a background,
    \begin{equation}
        n(\Delta) = n_0 - \frac{A}{2 \pi}\frac{\gamma}{(\Delta-\Delta_c)^2 + (\gamma/2)^2},
    \end{equation}
    where $n_0$ is the background dissociated atom number, $A$ is the area, $\gamma$ is the full width at half maximum, and $\Delta_c$ is the center frequency. We typically operate with 1 kHz peak widths and can measure peak position to $\sim$$100$ Hz.

    Since precise determination of transition frequencies is not the main purpose of this paper, we did not characterize the Stark shifts experimentally. Instead, we calculate a conservative upper bound on the uncertainty of the binding energy by combining estimated lattice and Raman laser Stark shifts.

    Using the Stark shift measured during magic wavelength determination, we fit a linear slope to Stark shift vs. lattice frequency near the operational magic wavelength. We then use this slope to convert the wavemeter-limited uncertainty of the lattice wavelength to a Stark shift, and take this Stark shift as our lattice contribution to the uncertainty of the binding energy.

    Using measured laser power and waist, as well as \emph{ab initio} polarizabilities calculated using the sum-over-states approach \cite{Bonin1997}, we calculate the Raman Stark shifts,
    \begin{equation}
        \Delta f_{\text {clock }}=\frac{I_R}{2 h \epsilon_0 c}\left[\alpha_0\left(\lambda_R\right)-\alpha_{62}\left(\lambda_R\right)\right],
    \end{equation}
    where $I_p$ is the intensity of each Raman laser, $\alpha$ is polarizability for each vibrational state, and $\lambda$ is the wavelength. We note that contributions from the Raman lasers have opposite signs~\cite{Kondov2019a}. We assign an additional conservative value of 50$\%$ to the \emph{ab initio} polarizabilities, significantly larger than the discrepancy observed in comparison with measured polarizability ratios \cite{Leung2023}.

    After estimating the lattice and Raman Stark shifts individually, we combine them to get total uncertainty on binding energy position. We find that the lattice Stark shift is about an order of magnitude greater than Raman Stark shift.

    \section*{S3. Finite sample temperature}
    Our experiment relies on Stark-induced shifts to molecular clock lines. Here we estimate the effect of finite sample temperature on the determination of differential polarizabilities from observed shifts.

    In the absence of the Stark laser the molecules, whether in their initial ($v$), or target ($v'$) vibrational states are trapped in the same magic-wavelength lattice potential. For a single lattice site this may be approximated by a harmonic trap potential:
    \begin{equation}
        V(x, y, z) =
        \frac{1}{2} M \omega_r^2 \left( x^2 + y^2 \right)
        + \frac{1}{2} M \omega_z^2 z^2.
    \end{equation}
    Here $M$ is the mass of the molecule and $\omega_{r, z}$ are the radial ($r$) and axial ($z$) trapping frequencies.

    We induce an ac Stark shift on the molecular clock $v\leftrightarrow v'$ transition by adding an extra collimated laser coaligned with the lattice which gives rise to an extra potential,
    \begin{equation}
        W(x, y, z) =
        \frac{1}{2} M \Omega_{v, v'}^2 \left(x^2 + y^2 \right) - U_{v,v'},
    \end{equation}
    where $\Omega_{v, v'}$ are the state-dependent radial trapping frequencies and $U_{v, v'}$ are the extra trapping depths. Both $\Omega_{v, v'}^2$ and $U_{v, v'}$ are directly proportional to the ac polarizabilites $\alpha_{v, v'}$ that we aim to measure. The increase in trap depth $U_{v, v'}$ leads to a temperature-independent line shift that is the basis for our experiment. However, the extra trapping frequency leads to a non-trivial temperature-dependent shift that we will evaluate here.

    The total trapping potential of the combined laser beams is
    \begin{equation}
        V_{v, v'} + W_{v, v'} =
        \frac{1}{2} M \left(\omega_r^2 + \Omega_{v, v'}^2 \right)
        \left(x^2 + y^2 \right)
        + \frac{1}{2} M \omega_z^2 z^2
        - U_{v, v'}.
    \end{equation}
    This is equivalent to a three-dimensional harmonic oscillator with state-dependent trapping frequencies. As carrier transitions preserve the motional quantum numbers, the total shift may be evaluated as a difference of the quantum thermal averages of the trapping hamiltonians $H_{v, v'} = T + V + W_{v, v'}$:
    \begin{eqnarray}
        \big<\delta E \big> & = & \big <H_{v'}\big> - \big<H_{v}\big> \nonumber \\
        & = & - \Delta U + \hbar \Delta\omega \left[\big<n_x \big> + \big<n_y\big> + 1\right],
    \end{eqnarray}
    where $\Delta U = U_{v'} - U_v$ and the change in radial trapping frequency is
    \begin{equation}
        \Delta \omega = \sqrt{\omega_r^2 + \Omega_{v'}^2} - \sqrt{\omega_r^2 + \Omega_{v'}^2 - \Delta(\Omega_{v,v'}^2)}.
    \end{equation}
    For us the transition-dependent term $\Delta (\Omega_{v,v'}^2) = \Omega_{v'}^2 - \Omega_{v}^2$ is on the whole substantially smaller than either of the trapping frequencies $\Omega_v^2$ or $\omega_r^2$, hence we can expand $\Delta\omega$ as
    \begin{equation}
        \Delta\omega \approx \sqrt{\omega_r^2 + \Omega_{v'}^2}
        \left(
        \frac{1}{2}\frac{\Delta(\Omega^2)}{ \omega_r^2 + \Omega_{v'}^2}
        - \frac{1}{8}\left(\frac{\Delta(\Omega^2)}{ \omega_r^2 + \Omega_{v'}^2}\right)^2
        \right).
        \label{eq:expansion}
    \end{equation}
    Importantly, the first term is linear in the measured differential polarizability as $\Delta(\Omega_{v, v'}^2)$ is directly proportional to $\Delta\alpha_{v,v'}$.

    The mean vibrational quantum numbers for radial motion can be evaluated by averaging over the grand canonical ensemble:
    \begin{eqnarray}
        \big<n_{x,y} \big> = \frac{1}{Z}\sum_{n=1}^\infty e^{\frac{-E_n(x,y)}{k_B T}} \approx \frac{k_B T}{\hbar \omega_r},\\
    \end{eqnarray}
    where we used the partition function \cite{LandauStatisticalPhysics}
    \begin{equation}
        Z = {\rm Tr}(e^{-H_{v'}/k_B T}) =  \frac{1}{2}{\rm csch}(\hbar\omega/2k_B T).
    \end{equation}
    Finally, the total thermally averaged shift to the line is
    \begin{equation}
        \big<\delta E \big> = -\Delta U + k_B T \frac{\Delta \omega}{\omega_r}.
    \end{equation}
    The first term is the temperature-independent ac Stark shift. The second term is a temperature-dependent correction.

    In our experiment the incoming lattice beam has a power of $P_l = 0.27$~W and a waist of $w_l = 36\,\mu{\rm m}$. For all the measured transitions the wavelength of the lattice is chosen to achieve a magic condition. This means that the polarizability at the lattice wavelength for both the initial $v'$ and target $v$ molecular states is the same and can be modeled as twice the atomic polarizability. For the magic wavelengths ranging from $\lambda_{\rm magic} = 996.4379$~nm to $\lambda_{\rm magic} = 1016.9714$~nm the atomic polarizabilities range from $\alpha_{\rm magic} = 250.2$~a.u. to $\alpha_{\rm magic} = 247.6$~a.u., respectively. This corresponds to total atomic trap depths
    \begin{equation}
        U_l = 4 \alpha P_l / (\pi w_l^2 c \epsilon_0)
    \end{equation}
    between $622\,{\rm kHz}\times h$ and $616\,{\rm kHz}\times h$ (approximately $30\,\mu{\rm K})$. The factor of four stems from constructive interference between the incident and reflected lattice beams. Conversely, the radial trapping frequencies
    \begin{equation}
        \omega_r = \frac{2}{w_l}\sqrt{U_l / M}
    \end{equation}
    of $\omega_r = 2\pi \times 469.9$~Hz to $2\pi \times 467.5$~Hz. The molecular sample temperature is estimated at $5\,\mu{\rm K}$. For weakly-bound molecules the trap depth is twice that for atoms (because the polarizability is that of two atoms), however, the trapping frequencies are the same for atoms and molecules, as the extra trap depth cancels out with the twice larger mass $M$ of the molecule.

    The extra Stark shift laser has a wavelength of $\lambda = 1950\,{\rm nm}$ and a maximum power of $P = 1.7\,{\rm W}$ at a waist of $w = 125.9\,\mu{\rm m}$. This provides an extra trap depth
    \begin{equation}
        U_v = \alpha_{x,v} P / (\pi w^2 c \epsilon_0)
    \end{equation}
    between $67.1\,{\rm kHz}\times h$ and $79.4\,{\rm kHz}\times h$ \emph{per atom in the molecule} and an extra radial confinement that varies from $\Omega_v = 2\pi \times 44.1$~Hz to $\Omega_v = 2\pi \times 48.0$~Hz. The polarizability per atom varies between $\alpha_x = 210.1$~a.u. for the most weakly bound state and $\alpha_x = 248.5$~a.u. for the rovibrational ground state.

    The temperature-independent shift $\Delta U$ of up to $12.3\,{\rm kHz}\times h$ by far outweighs the temperature-dependent term. Note that the total shift for the diatomic molecule is $2\Delta U$ and it therefore reaches $24.6\,{\rm kHz}\times h$. The extra temperature-dependent term stems from the change in the total radial confinement of the effective trap created by the lattice and ac Stark laser. The contribution to radial trapping from the ac Stark laser is an order of magnitude smaller than the baseline provided by the lattice laser.
    The figure of merit is the difference in the extra confinement between different vibrational states as compared to the lattice radial frequency.
    We find $\Delta \omega$ to vary between $2\pi \times 6.4 \times 10^{-4}$~Hz for a $v = 61\leftrightarrow v' = 62$ transition and $2\pi \times 0.38$~Hz for a $v = 0\leftrightarrow v' = 62$ line. The total temperature-dependent shift, $k_B T (\Delta \omega / \omega_r)$ is consistently below $0.7\%$ of the temperature-independent shift, at most $84.3~{\rm Hz} \times h$ for the $v = 0\leftrightarrow v' = 62$ transition. This is already significantly smaller than our experimental error bars.

    We also point out that most of the thermal shift is, just like $\Delta U$, directly proportional to the differential polarizability we aim to measure. In fact, the linear term in the Taylor expansion for $\Delta \omega$, Eq. (\ref{eq:expansion}) overestimates the real value by at most $0.4\%$ making the nonlinear systematic negligible for this work.

    Another possible source of systematic error is the variation in magic wavelength and the corresponding lattice radial confinenement between the different molecular lines. We find that this variation contributes at most a $1.2\%$ relative uncertainty to the temperature-dependent shift. Again, for us this contribution is two order of magnitude smaller than our experimental uncertainty and therefore negligible.

    \section*{S4. Uncertainty of the theoretical polarizabilities due to empirical potential}

    \begin{figure}
        \includegraphics[width=\columnwidth]{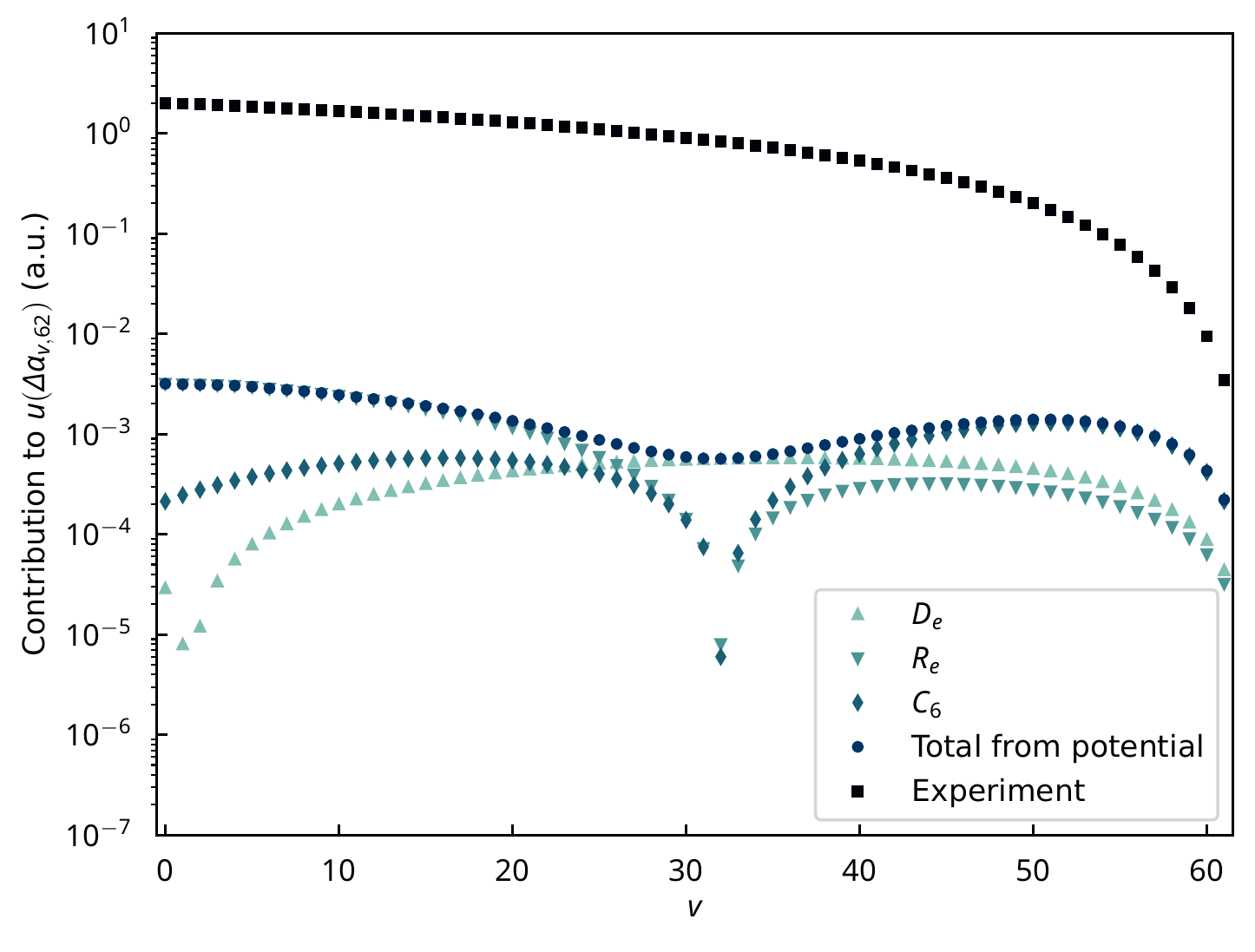}
        \caption{Contributions to the uncertainty of the theoretical polarizabilities due to the use of empirical potential from Ref.~\cite{Stein2010}. For all investigated transitions the contributions are all at least one order (typically more than two) of magnitude smaller than the error bar assigned to our theoretical model through comparison with experiment.}
        \label{fig:potentialError}
    \end{figure}

    To calculate the blackbody radiation shifts we needed the vibrational wavefunctions for all nonrotating vibrational states of strontium molecules in their electronic ground state. These were obtained by solving the radial Schr\"odinger equation using an accurate potential obtained empirically from Fourier transform spectroscopy~\cite{Stein2010}. The potential was provided in two versions: one in terms of a piecewise function and as a Morse/Long-Range (MLR) fit~\cite{LeRoy2009}. We used the latter.

    In the paper we estimated the uncertainty of the theoretical model by comparing theoretical ac polarizabilities to experimental data and concluded that model is accurate to within 2.6\%. We expect that most of this error bar is coming from the combination of the limited accuracy of the \emph{ab initio} polarizabilities and the experimental accuracy. Here, we additionally look at the uncertainty of the theoretical model stemming from the use of an empirical potential~\cite{Stein2010}. To estimate the sensitivity of the theoretical polarizabilities to the experimental uncertainty we vary three of the most important parameters of the potential -- dissociation energy $D_e$, equilibrium distance $R_e$ and the leading van der Waals coefficient $C_6$ and rerun our calculation. The parameters $R_e = 4.6720(1)\,{\rm \AA}$ and $D_e = 1081.64(2)\,{\rm cm}^{-1}$~\cite{Stein2010} are varied within their stated experimental uncertaintes whereas $C_6$ was varied such that the (well known) position of the near-threshold $v=62$ bound state at $-137\,{\rm MHz}$ shifted by at most 1~MHz.

    The contributions to due to $D_e$, $R_e$ and $C_6$ are shown in Fig.~\ref{fig:potentialError}. The variation of each parameter influences the predicted polarizabilities in a distinct manner. As the polarizability depends chiefly on the mean internuclear distance of a given vibrational level, scaling the potential depth $D_e$, for example, has little influence on the polarizability of deeply bound states. On the other hand, these states are naturally more sensitive to varying the equilibrium distance $R_e$. Lastly, weakly bound states are the most sensitive to the variation of the long-range van der Waals interaction coefficient, $C_6$. Nevertheless, we find that all of these error contributions are at least one order of magnitude smaller than the uncertainty we assigned to the model via direct measurements of ac Stark shifts and for our purposes are negligible.

    \section*{S5. The Planck integrals}

    Our calculation of the blackbody radiation shift relies on expanding the differential polarizability of a transition in terms of a series of Cauchy coefficients. Averaging each contribution to the polarizability over the Planck distribution involves calculating integrals of the following type:
    \begin{equation}
        c_n = \int_0^\infty \frac{u^{3+n}}{\exp(u) - 1} du = {\rm Li}_{n+4}(1) \Gamma (n+4)
    \end{equation}
    for even $n$. Here ${\rm Li}_{s}(z)$ is the polylogarithm function of order~$s$,
    \begin{equation}
        {\rm Li}_{s}(z) = \sum_{k=1}^\infty \frac{z^k}{k^s},
    \end{equation}
    and $\Gamma(x)$ is Euler's gamma function. While for our purposes it was enough to cut the series off at $n=4$, in the future higher orders might be needed. For future reference, here we list the first eight integrals:
\begin{eqnarray}
    c_{ 0 } & = &  \frac{\pi^{4}}{15}  \approx  6.49393940226683 \ldots \nonumber \\
    c_{ 2 } & = &  \frac{8 \pi^{6}}{63}  \approx  122.081167438134 \ldots \nonumber \\
    c_{ 4 } & = &  \frac{8 \pi^{8}}{15}  \approx  5060.54987523764 \ldots \nonumber \\
    c_{ 6 } & = &  \frac{128 \pi^{10}}{33}  \approx  363240.911422383 \ldots \nonumber \\
    c_{ 8 } & = &  \frac{176896 \pi^{12}}{4095}  \approx  39926622.9877311 \ldots \nonumber \\
    c_{ 10 } & = &  \frac{2048 \pi^{14}}{3}  \approx  6227402193.41097
    \ldots \nonumber \\
    c_{ 12 } & = &  \frac{3703808 \pi^{16}}{255}  \approx  1307694352218.91 \ldots \nonumber \\
    c_{ 14 } & = &  \frac{1437433856 \pi^{18}}{3591}  \approx  355688785859224 \, .
\end{eqnarray}

\bibliography{library}